\documentclass[preprint,floatfix
,aps,prd,superscriptaddress]{revtex4-2}

\usepackage[utf8]{inputenc}
\usepackage{tikz}
\usepackage{amssymb,amsmath,mathtools,comment,hyperref}
\usepackage{appendix}
\usepackage[utf8]{inputenc}
\usepackage{physics}

\usepackage{array, makecell}
\usepackage{cellspace} %
\usepackage{extarrows}
\usepackage{relsize}

\usepackage{bm}
\usepackage{mathrsfs}
\usepackage{xcolor}

\usepackage{dcolumn}

\begin{document}

\title{
Conformal quantum mechanics of causal diamonds: 
\\
Time evolution, thermality, and instability 
\\ via path integral functionals
}

\author{H. E. Camblong}
\affiliation{Department of Physics and Astronomy, University of San Francisco, San Francisco, California 94117-1080, USA}
\author{A. Chakraborty}
\affiliation{Department of Physics, University of Houston, Houston, Texas 77024-5005, USA}
\affiliation{Institute for Quantum Computing, Department of Physics and Astronomy, 
University of Waterloo, Waterloo, Ontario N2L 3G1, Canada\looseness=-1}

\author{P. Lopez-Duque}
\affiliation{Department of Physics, University of Houston, Houston, Texas 77024-5005, USA}
\author{C. R. Ord\'o\~nez}
\affiliation{Department of Physics, University of Houston, Houston, Texas 77024-5005, USA}

\date{December 17, 2024}

\begin{abstract}
An observer with a finite lifetime $\mathcal{T}$ perceives the Minkowski vacuum 
as a thermal state at temperature $T_D = 2 \hbar/(\pi \mathcal{T})$, as a result of 
being constrained to a double-coned-shaped region known as a causal diamond.  
 In this paper, we explore the emergence of thermality in causal diamonds due to the
 role played by the symmetries of conformal quantum mechanics (CQM) as a (0+1)-dimensional conformal field theory,
within the de Alfaro-Fubini-Furlan model and generalizations. 
In this context, the hyperbolic operator $S$ of the SO(2,1) symmetry of CQM: 
(i)  is the generator of the time evolution of a diamond observer;
(ii) its dynamical behavior leads to the predicted thermal nature;
and (iii) its associated quantum instability has a Lyapunov exponent $\lambda_L = \pi T_D/\hbar$, which is half the upper saturation bound of the information scrambling rate.
Our approach is based on a comprehensive framework of path-integral representations of the CQM generators in canonical and microcanonical forms, supplemented by semiclassical arguments. The 
  properties of the operator $S$ are studied with emphasis on an operator duality with the corresponding elliptic operator $R$, using a representation in terms of an effective scale-invariant inverse square potential combined with inverted and ordinary harmonic oscillator potentials.

\end{abstract}


\maketitle
\newpage

\section{Introduction}
\label{sec:introduction}

This paper addresses the thermal nature of causal diamonds in flat spacetime.
In principle, the thermality of the Minkowski vacuum corresponds to the detection of particles 
with a thermal distribution by a finite-lifetime observer.
The theory is defined in terms of the action of a conformal quantum mechanics (CQM) model
as a $(0+1)$-dimensional field theory, with focus on a single-component field.  The implementation 
of the formalism is performed via the properties of the SO(2,1) conformal symmetry generators.
Our main results, which are fully quantum-mechanical and based on the propagator and Green's function functionals of CQM, corroborate and complement similar findings from other methods.

\subsection{Horizons and thermal behavior in spacetime}

A generic prediction of relativistic quantum field theory in the presence of horizons is thermal behavior~\cite{birrell-davies}.
This was discovered by Hawking
in a series of seminal papers~\cite{hawking1,hawking2,hawking3} showing 
that black holes emit thermal radiation with a temperature $T_H = \hbar \kappa/2\pi$, where $\kappa$ is the surface gravity of the black hole. 
(In this paper, we will use units with $ k_B = c = 1$.)
In essence, the event horizon restricts causal access, making the quantum density matrix be a mixed, 
thermal state~\cite{birrell-davies}.
This prediction, which has been confirmed by a variety of related theoretical treatments, 
is expected to be a basic test of any candidate theory of quantum gravity~\cite{wald-rev,page-rev}.
Black hole radiance also confirms Bekenstein's identification of the black hole entropy $S_{\rm BH} $
with the area $A_{\rm BH}$ of the black hole event horizon~\cite{bekenstein1972, bekenstein1973},
uniquely fixing its value as
$S_{\rm BH} 
= A_{\rm BH}/4l_{P}^2$, where $l_{P}=\sqrt{\hbar G}$ is the Planck length;
in addition, it gives
a self-contained framework for black hole thermodynamics~\cite{hawking3,wald-rev,page-rev}.

\begin{figure}[h]
    \centering
    \includegraphics[width=0.55\linewidth]{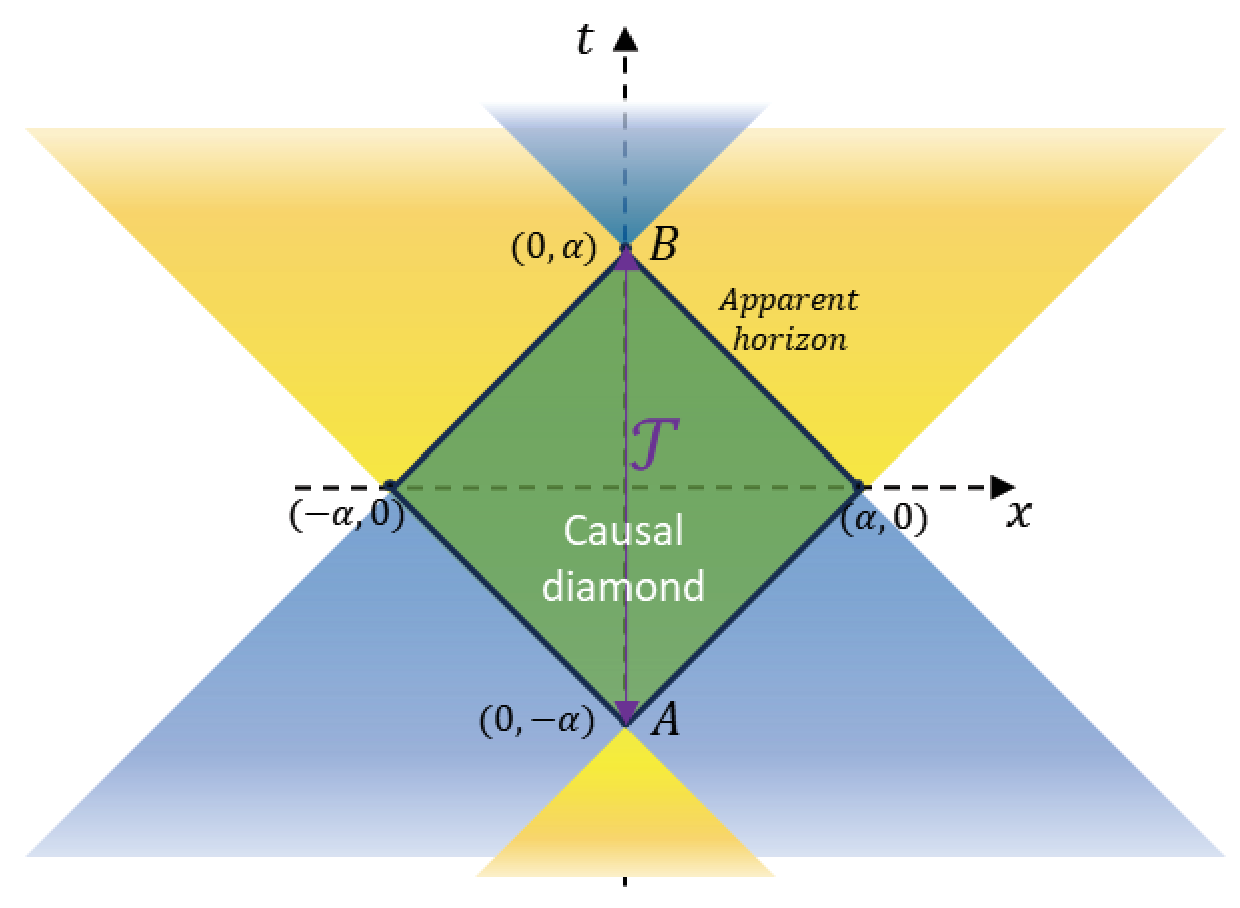}
    \caption{The green region denotes the causal diamond of size $2\alpha$ (``radius'' $\alpha$),
    which is the spacetime region accessible to an observer with a finite lifetime $\tau = 2 \alpha$. 
    The diamond's boundaries are causal horizons.}
    \label{fig:causal-diamond}
\end{figure}

In a parallel development, while black-hole thermal effects and Hawking radiation manifest as curved-spacetime realizations, similar \textit{thermal properties are also found in the presence of horizons that restrict causal access within flat, Minkowski spacetime\/.} Most importantly, 
the Fulling-Davies-Unruh effect~\cite{Fulling-1973_CFQ,Davies-1975_Unruh,unruh-notes,ufd}
shows that observers with an acceleration $a$ detect thermal particles with temperature
$T_U = \hbar a/2\pi$ in the Minkowski vacuum due to the presence of a Rindler horizon. In this setting,
the horizon limits causal access such that that a pure state defined over the whole Minkowski spacetime appears as a mixed, thermal state to an accelerated observer. Remarkably, thermal effects are also expected for finite-lifetime observers because their causal access is limited to a double-coned-shaped region of Minkowski spacetime, which is the intersection of the future and past light cones between an initial and a final time, respectively (see Fig.~\ref{fig:causal-diamond}). 
This restricted region, known as the causal diamond, makes a finite-lifetime observer, with lifetime 
$\mathcal{T} = 2\alpha$ (where $2\alpha$ is the size of the diamond), 
perceive the Minkowski vacuum as thermal,
as shown in Ref.~\cite{martinetti-1}, 
extending the work of Refs.~\cite{Bisognano-Wichmann, Hislop-Longo,Haag_LQP}.
Moreover, this thermal background consists of excitations with particles at the temperature
\begin{equation}
T_D = \frac{\hbar}{\pi \alpha}
\; ,
\label{eq:diamond-temperature}
\end{equation} 
which can be probed with an energy-scaled detector,
as proposed in Ref.~\cite{Su2016SpacetimeDiamonds} and further developed in Ref.~\cite{Chakraborty:2022oqs}
with an open quantum systems approach.
Other findings on the thermality of causal diamonds have confirmed this picture~\cite{martinetti-2, casini:2011,Ida-etal_2013,DeLorenzo_light-cone,jacobson2019gravitational,Foo2020GeneratingMirror,Tian_DS, Good-etal_DS,DeLorenzo_BH}.
It is noteworthy that generic thermal effects driven by the presence of horizons in flat spacetime provide insightful conceptual and mathematical laboratories to understand quantum aspects of spacetime, including strong-gravity effects and black-hole radiation; and they may eventually lead to simpler experimental realizations of similar physics in the laboratory. An example of the nontrivial thermal physics of causal diamonds is provided by the recent analysis of entanglement of a bipartite system~\cite{camblong-ent-deg-2024}. Specifically, it was found that if the system starts in a maximally entangled state prepared from an inertial perspective, entanglement is degraded due to the presence of the diamond's causal horizons. 
These results give a physical realization of the conjecture that causal diamonds appear to be at the heart of the thermodynamic properties of spacetime and gravity~\cite{jacobson2019gravitational}.

\subsection{Connections of conformal quantum mechanics (CQM) with the physics of causal diamonds and other spacetime realizations}

In this paper, we further develop the deep connections between causal diamonds and CQM,
and their crucial role supporting the thermal character of the quantum state detected by finite-lifetime observers.
Thus, we add to the extensive evidence that
\textit{conformal symmetry generically and conformal quantum mechanics specifically play central roles in all thermodynamic effects in spacetime.\/} 
This remarkable insight has been revealed in a variety of independent approaches for black holes, including string theory~\cite{strominger-vafa,strominger},
horizon 2D conformal symmetries~\cite{carlip1,carlip2}, 
 the AdS/CFT correspondence~\cite{ryu-takyanagi}, 
 the Kerr/CFT correspondence~\cite{stromingerkerr}--\cite{strominger10},
and the tunneling formalism~\cite{srinivasan-complex}, among others---and this is still a fertile ground for fundamental approaches to gravitational theory~\cite{almheiri-rev}. In this setting,
CQM, with an SO(2,1) symmetry,
has been used as a near-horizon model for generic questions of near-horizon black hole physics: basic black hole thermodynamics, including the origin of the black hole entropy~\cite{nhcamblong, nhcamblong-sc, nhcamblong-heat-kernel, nhcamblong-conformal-tightness}, and black-hole acceleration radiation~\cite{acceler-rad-Schwarzschild, acceler-rad-Kerr,acceler-rad-Qopt-1, acceler-rad-Qopt-2}; this adds to partial evidence found in other references regarding the near-horizon role of the inverse square 
 potential~\cite{srinivasan-complex,Padmanabhan_BH-ISP,guptasen}.
 Independently, the relation between black-hole radiation and the dilatation operator $D$ of CQM has also been discussed in the recent literature~\cite{bala-IHO-review} (within the scope of the inverted harmonic oscillator). 
  
These miscellaneous connections between spacetime properties and conformal quantum mechanics
are of special interest in light of recent findings in quantum information theory and quantum chaos.
Specifically, in a recent series of insightful 
papers~\cite{majhi-nh-chaos,majhi-nh-thermality1,majhi-nh-thermality2,majhi-thermality-Kerr,majhi-nh-thermality-ansatz,majhi-nh-thermality-PI,betzios}, it was shown that the quantum chaotic nature of a black hole can be probed using the scale-invariant $xp$-type potential observed by an infalling particle near the event horizon. This $xp$-type potential corresponds to the dilatation operator of CQM and has been discussed in great detail in the context of quantum aspects of chaos \cite{berry-keating,sierra}.
More generally,
black holes appear to be the fastest information scramblers available in nature, 
given that the only information coming out of a black hole is thermal in nature~\cite{lenny2008}. 
This insight led to the conjecture~\cite{MSS-bound} that there exists a bound on the rate of growth of chaos in thermal quantum systems, which can be described as the information scrambling rate quantified by the out-of-time-order correlation (OTOC) function. This diagnostic tool for quantum chaos has become of widespread use in a variety of physical systems~\cite{OTOC1,OTOC2,OTOC3,OTOC4,OTOC5}; 
 the OTOC can grow exponentially in a quantum chaotic system, 
 with the information scrambling rate (quantum Lyapunov exponent) $\lambda_L$ having the upper bound
 \begin{equation}
 \lambda_L \leq \frac{2\pi T}{ \hbar}
 \; ,
 \label{eq:Qscrambling_bound}
 \end{equation}
  where $T$ is the temperature of the system. For a black hole, the growth rate of information scrambling 
  saturates the upper bound, with $\lambda_L = 2\pi T_H/\hbar$, where $T_H$ is the Hawking temperature. 
  Hence, black holes can be considered as maximally chaotic quantum systems. Moreover, this led to the conjecture that the bound saturation of a system is the signature that it is holographic dual to a 
  black hole~\cite{OTOC3,Shenker-Stanford_BHchaos-stringy}, 
  with an example provided by the Sachdev-Ye-Kitaev (SYK) model~\cite{SYK-model,syk-maldacena,chaos-ads2,Qinfo-review_LewisSwan}. 

Our construction, based on CQM, is focused on the role played by the generator of time evolution 
of a finite-lifetime observer, which is the hyperbolic operator $S$~\cite{arzano,arzano2}. 
This is one of the noncompact generators of the SO(2,1) group, along with the elliptic ($R$) and dilatation  ($D$) generators, as first derived in the seminal paper of the dAFF model of de Alfaro, Fubini, and Furlan~\cite{dAFF}. 
The causal-diamond temperature~(\ref{eq:diamond-temperature})
 can be explained using the ${\rm CFT_1}$ algebra of the AdS/CFT correspondence
 from a CQM group-theoretic perspective~\cite{arzano,arzano2}.
However, it would be desirable to explore
an independent, more physical approach to the thermal nature of causal diamonds within the dAFF model.
Here, we fill this gap using 
\textit{the operator $S$ as an effective Hamiltonian, represented in terms of an inverse-square potential 
and an inverted harmonic oscillator,
 and governing the causal-diamond dynamics.\/}
 Specifically:
 \begin{itemize} 
\item
 We establish an operator duality relating the hyperbolic and elliptic families of the symmetry generators;
 this amounts to a duality between the operators $S$ and $R$. 
 In the effective potential representation, this corresponds to an inverse square potential coupled to 
 inverted and ordinary harmonic oscillators, respectively. 
 \item
We determine the diamond temperature $T_D$ of Eq.~(\ref{eq:diamond-temperature})
associated with the evolution of the operator $S$,
with a correspondence via its dual operator $R$.
\item
By a direct use of path-integral methods with the known functionals of the CQM generators~\cite{PI-CQM},
we completely characterize the thermal behavior of causal diamonds with the temperature $T_D$.
\item
We find evidence that the thermality of causal diamonds
is related to the quantum instability of the evolution driven by the $S$ operator,
which can be identified in the semiclassical regime.
This corroborates and expands
the insightful findings of 
      Refs.~\cite{majhi-nh-chaos,majhi-nh-thermality1,majhi-nh-thermality2,majhi-thermality-Kerr,majhi-nh-thermality-ansatz,majhi-nh-thermality-PI} 
      for black holes, within the simpler setting of causal diamonds in flat spacetime. Most importantly, 
     this framework provides \textit{insight into the quantum scrambling inequality\/}
      of Eq.~(\ref{eq:Qscrambling_bound}), 
     which causal diamonds do satisfy with
     $\lambda_{L} = 1/\alpha $ equal to one half the upper bound
     $2\pi T_{D} /\hbar = 2/\alpha $.

\end{itemize}

\subsection{Outline}

This paper is organized as follows.
In Sec.~\ref{sec:CQM-review}, 
we review the definition and properties of CQM, with emphasis on its symmetry generators
and the specifics of the inverse-square potential dAFF model. 
 In Sec.~\ref {sec:causal-diamonds_CQM},
 we explain how the time evolution in causal diamonds is driven by the 
 hyperbolic generator $S$.
  The main results of this work are presented in
 Sec.~\ref{sec:semiclassical_GF-thermality}, where we show how this CQM time evolution
 leads to {\em a thermal behavior at the diamond temperature $T_D$, within a framework of path-integral relations\/} 
 that explicitly relies on CQM operator dualities.
In Sec.~\ref{sec:semiclassical_instability-thermality}, 
we show the emergence of a dynamical instability within a semiclassical approach due to 
 the same $S$-driven time evolution in phase space; and
 we explore additional aspects of the semiclassical behavior,
 including thermality.
 In Sec.~\ref{sec:discussion},
 we offer a set of concluding remarks, with
 additional insight into the physical meaning and prospects of these findings.
   The appendices summarize: 
   (i) the definitions and properties of path integrals needed in the main text; 
  (ii) the technical details of the path-integral representations of the CQM generators, 
 with relevant properties for the thermality of this system; and
 (iii) the semiclassical and fully quantum-mechanical expressions for the density of states.

\section{Conformal Quantum Mechanics (CQM): Symmetries and Dynamics}
\label{sec:CQM-review}

This section is a brief 
review of the definition and properties of conformal quantum mechanics (CQM) tailored to the central goal of our paper:
a thorough description of the time-evolution dynamics of causal diamonds 
and thermality driven by the CQM generator $S$.
In this section, and in most of the main 
body of this work until Sec.~\ref{sec:semiclassical_GF-thermality}---but not in the appendices---we 
will use units with $\hbar =1$.
The dAAF model of CQM was originally developed in the seminal work by
 de Alfaro, Fubini, and Furlan~\cite{dAFF}, and has received considerable attention over the decades. 
 For the discussion below,
 we will use the notation and emphasis of our recent study of CQM with path integrals~\cite{PI-CQM}.

\subsection{CQM action and symmetries: Generators of \boldmath{${so}$}(2,1) algebra}
\label{subsec:gen-CQM}

The original dAAF model~\cite{dAFF} 
is a $(0+1)$-dimensional conformal field theory, with the action given by
\begin{equation}
    S_{\!_{\rm CQM}} [Q(t)] = \int dt\;\left(\frac{1}{2}\dot{Q}^2 - \frac{g}{2Q^2}\right)
    \;,
    \label{eq:CQM_action}
\end{equation}
where $Q\equiv Q(t)$ is a generalized position coordinate, $\dot{Q}$ is the derivative with respect to time $t$ 
(which is identical to the conjugate momentum), and $g>0$ is a dimensionless coupling constant. 
The term $g/2Q^2$ is the well-known inverse-square potential that finds a wide range of 
applications in molecular physics~\cite{Qanomaly-molecular,Qanomaly-molecular-EFT,Qanomaly-molecular-to-BH}, nanophysics~\cite{CQM-renormalization-PLA}, 
nuclear and particle physics~\cite{CQM-renormalization-PLA}, 
and black-hole thermodynamics~\cite{nhcamblong, nhcamblong-sc, nhcamblong-heat-kernel, nhcamblong-conformal-tightness, Qanomaly-molecular-to-BH,acceler-rad-Schwarzschild, acceler-rad-Kerr}.
In what follows, we will summarize the symmetry framework introduced
in Ref.~\cite{dAFF},
 and more generally analyzed within a path-integral approach in Ref.~\cite{PI-CQM}.

The dAFF model, as defined by the action~(\ref{eq:CQM_action}), is invariant under the transformation
\begin{align}
    t' &= \frac{a t + b}{c t + d}\;, \label{eq:sl2r_transform-a}\\
    Q'(t') &= (ct+d)^{-1} Q(t)
     \label{eq:sl2r_transform-b}
    \;,
\end{align}
where $a, b, c, d$ are real numbers subject to the constraint $ad-bc = 1$.
The transformation of the time parameter can be represented by a matrix
\begin{equation}
    \begingroup 
    \setlength\arraycolsep{6pt}
    M = \Bigg( \; \begin{matrix}
     a & b\\
     c & d
    \end{matrix}\; \Bigg)
    \;,
    \endgroup
\end{equation}
which belongs to the SL(2,$\mathbb{R}$) group and enforces the group properties of 
 Eqs.~(\ref{eq:sl2r_transform-a})--(\ref{eq:sl2r_transform-b}),
 with the following three group generators, which are written in terms of canonical variables $(Q,P)$,
 with canonical momentum $P = \dot{Q} $, according to the action of Eq.~(\ref{eq:CQM_action}).
 \begin{itemize}
    \item Time-translation generator: 
\begin{equation}
  H = \frac{1}{2} \, P^2 + \frac{g}{2Q^2}
  \; .
    \label{eq:T-translation-generator}
  \end{equation}
    \item Dilatation generator: 
  \begin{equation}
   D = tH - \frac{1}{4}(Q P+ P Q)
   \; .
     \label{eq:dilation-generator}
   \end{equation}
    \item Special conformal transformation generator: 
\begin{equation}
\displaystyle K = t^2 H - \frac{1}{2}t(Q P+PQ) + \frac{1}{2}Q^2
  \label{eq:SCT-generator}
  \; .
  \end{equation}
\end{itemize}
These generators are elements of the $sl(2,\mathbb{R})$ Lie algebra with commutators
 \begin{equation}
    [D,H] = - iH
    \; , \qquad\qquad 
    [D,K ] = iK
    \; , \qquad\qquad 
    [H,K] = 2iD
    \; .
    \label{eq:sl2R-algebra}
     \end{equation}
 In addition, in the Cartan-Weyl basis~\cite{wyb:74},
 the generators $H,\, K$, and $D$ are replaced by the linear combinations
\begin{align}
    \hat{R} &= \frac{1}{2}\left(\alpha H + \frac{1}{\alpha} \, K \right)\;,\\
    S' &= - \hat{S} = \frac{1}{2}\left(\alpha H - \frac{1}{\alpha} \, K \right) \; ,
      \label{eq:hat-R-S'-def}
\end{align}
where $\hat{R} \equiv R_{dAFF}$ and 
$\hat{S} \equiv S_{dAFF}$ are the conventional choices used in the original dAFF model~\cite{dAFF}. 
The dissimilar physical dimensions of 
$H$ (inverse time) and $K$ (time) in the Cartan-Weyl linear combinations require 
the existence of an arbitrary parameter $\alpha$ with dimensions of time. 
The physical meaning of this parameter $\alpha$ can be further developed in terms of the time evolution within the causal diamond, as will be discussed in the next section. 
Then, the operators $\hat{R},\hat{S}= - S'$, and $D$ satisfy the $so(2,1)$ algebra,
\begin{equation}
    [D,\hat{R}] = i \hat{S} \;,\qquad\qquad [\hat{R},\hat{S}] = i D\;,\qquad\qquad [D, \hat{S}] = i \hat{R} 
    \; ,
    \label{eq:SO-21-algebra}
\end{equation}
which is homomorphic to the 
 $sl(2,\mathbb{R})$ algebra~(\ref{eq:sl2R-algebra}) of the original symmetry generators.
The $so(2,1)$ algebra structure
involves the compact operator $\hat{R}$ combined with the noncompact operators $\hat{S}$ and $D$. 
This implies that the time-evolution orbits of $\hat{R}$ are closed and periodic, whereas orbits of $\hat{S}$ and $D$ are unbounded~\cite{dAFF,PI-CQM}. 
Most importantly, the algebra of Eq.~(\ref{eq:SO-21-algebra}) identifies
 a {\em symmetry structure, described by the group SO(2,1), that fits into a general pattern
of symmetries of conformal field theory~\cite{CFT_DiFranceso}.\/}

It should be noted that the operator $S'\equiv -S_{dAFF}$ was introduced as an alternative to the original $S_{dAFF}$ in~\cite{PI-CQM} to get a more transparent physical interpretation of the corresponding effective potential (see Refs.~\cite{arzano} and \cite{PI-CQM}). 
In this article, we will rescale the Cartan-Weyl generators $\hat{R}$ and $S'= -\hat{S}$.
with a factor $2\alpha^{-1}$, i.e.,
\begin{align}
&
R = \frac{2}{\alpha} \hat{R} 
= H + \frac{1}{\alpha^2} \, K
 \;,
\label{eq: rescaled-R}
\\
  S = & \frac{2}{\alpha} {S'}   =-  \frac{2}{\alpha} \hat{S}  
  = H - \frac{1}{\alpha^2} \, K
  \label{eq: rescaled-S}
  \; ,
  \end{align}
  so that the kinetic terms in the resultant operators $R$ and $S$ have the standard 
form $\dot{Q}^2/2$, with the usual physical interpretation, just as in the original Hamiltonian $H$. 
This physically-motivated assignment automatically makes the corresponding effective time-evolution parameters
 $\tau$ have the correct physical time 
dimensions, with values that match the time $t$ of the original Hamiltonian $H$ 
as $\alpha \rightarrow \infty$~\cite{arzano}. 
Due to this rescaling, the corresponding commutators become:
$  [D,{R}] =- i {S} 
, \; [{R},{S}] = - 4 i D/\alpha^2\
, \; [D, {S}] = - i {R} $;
but all other characteristics remain the same,
including their time evolution and their path-integral functionals~\cite{PI-CQM}.
This conforms to the original results of Ref.~\cite{dAFF} 
and the more general expressions of Ref.~\cite{PI-CQM}.

Finally, one additional formal advantage of the rescaled choices of Eqs.~(\ref{eq: rescaled-R})-(\ref{eq: rescaled-S}) is that the analytic continuation 
$\alpha^{-1}  \longrightarrow -i {\alpha}^{-1} $
 changes the operator $R$ to $S$, as shown in Ref.~\cite{PI-CQM}. This important 
 connection will be further developed as an operator duality in Sec.~\ref{sec:semiclassical_GF-thermality},
 to display the emergence of thermal behavior.
  A more general analysis of the CQM operators follows next.

\subsection{CQM generalized Hamiltonians and dynamics}

The properties of CQM, including the uniqueness of the symmetry generators,
is based on enforcing the condition of conformal symmetry under arbitrary time transformations,
at the level of the action~(\ref{eq:CQM_action}).
Once the symmetry algebra is established, all the operators $R, S, D$ are on equal footing, and can be used 
as generators of effective-time translations.
This is a key concept introduced in the original dAFF model~\cite{dAFF}, leading to the 
generalized symmetry generators
\begin{equation}
    G = uH +vD + wK
    \; ,
\label{eq:generalized-generators}
\end{equation}
which can be interpreted as effective Hamiltonians of a redefined theory in terms of a 
corresponding dynamical time $\tau$. Moreover, the operators~(\ref{eq:generalized-generators}) 
can be completely characterized in terms of the discriminant 
\begin{equation}
\Delta = v^2 - 4uw
\; 
\label{eq:gen-generator_discriminant}
\end{equation}
that determines its nature:
rotations (elliptic type), for $\Delta <0$, which include $R$;
boosts (hyperbolic type), for $\Delta >0$, 
which include both $S$ and $D$;
 and parabolic (``lightlike'') operators, for $\Delta =0$,
 which include the original $H$ and $K$ generators.

Then, the resultant generalized dynamics governed by the effective Hamiltonian~(\ref{eq:generalized-generators}) as the 
time-translation generator involves an effective time $\tau$ given by
\begin{equation}
    d\tau = \frac{dt}{u+vt+wt^2}
    \; , \label{eq:tau-t-generic}
\end{equation}
where $t$ is the standard time corresponding to the original Hamiltonian $H$.
In addition, the associated field or coordinate variable $q(\tau)$ is given by
\begin{equation}
    q(\tau) = \frac{Q(t)}{|u+vt+wt^2|^{1/2}}
    \; ,
     \label{eq:q-Q-relation}
\end{equation}
with a corresponding canonical momentum
\begin{equation}
{p} =   \sigma \left| f_{G} \right|^{1/2} 
 \biggl(
{P} - \frac{\dot{ f_{G}} }{ 2 f_{G} } \,   {Q} 
 \biggr)
 \; ,
 \label{eq:momentum-p_q}
 \end{equation}
where~\cite{PI-CQM}
\begin{equation}
f_{G} (t) = u + vt + wt^{2} = \sigma \, \left| u + vt + wt^{2} \right|
\; ,
\label{eq:f(t)}
\end{equation}
and the sign $ \sigma \equiv \sigma_{G} = \mathrm{sgn} [f_{G}(t)] $ allows for
arbitrary linear combinations of the generalized generator~(\ref{eq:generalized-generators}).
This analysis, pioneered in Ref.~\cite{dAFF}, and further developed (for any number of dimensions)
in Ref.~\cite{PI-CQM}, shows that $G$ acts as a time-translation operator
 in the generalized Schr\"{o}dinger picture
\begin{equation}
    G\ket{\psi(\tau)} = i \frac{d}{d\tau} \ket{\psi(\tau)}
    \; ,
\end{equation}
with $\ket{\psi(\tau)} = e^{- i G (\tau-\tau_{0})} \ket{\psi(\tau_{0})}$.

The CQM generalized dynamics admits a Hamiltonian representation for each generalized generator $G$.
The conjugate generalized momentum, after applying the transformations of 
Eqs.~(\ref{eq:tau-t-generic})--(\ref{eq:q-Q-relation}), is given by Eq.~(\ref{eq:momentum-p_q}), which reduces to
 $p=\dot{q}$, where $\dot{q}(\tau)$ denotes the derivative with respect to its associated time $\tau$.
The corresponding effective Hamiltonian is
\begin{equation}
    \tilde{H}_{G}
       =
     \frac{1}{2}\, p^2 + \frac{g}{2q^2} +
     \frac{1}{2} \left(-\frac{\Delta}{4} \right)
     q^2
     \; ,
       \label{eq:Hamiltonians-for-G}
\end{equation}
(after removing the extra sign $ \sigma = \mathrm{sgn} [f_{G}(t)] $ and rescaling)~\cite{PI-CQM}.
With such Hamiltonian formulations, the 
evolution in phase space can be described with points $(q(\tau),p(\tau))$ flowing according to the 
effective times $\tau$, which are operator dependent, as shown in Eq.~(\ref{eq:tau-t-generic}).
In addition,
as a convenient effective-physics interpretation of the generators,
Eq.~(\ref{eq:Hamiltonians-for-G})
gives an exact analog of a quantum-mechanical problem with potential 
\begin{equation}
 \tilde{V} (q) =
\frac{1}{2} \, \frac{g}{q^{2}} +   \frac{1}{2} \omega^2 q^2 
\label{eq:potential_HO+ISP}
 \; ,
\end{equation}
yielding the original inverse square potential superimposed with a harmonic oscillator of squared ``frequency''
$\displaystyle
\omega^{2} = -\frac{\Delta}{4 }$
(allowing for both real and imaginary frequencies).

\begin{figure}[htb!]
       \centering
      \includegraphics[width=0.55\linewidth]{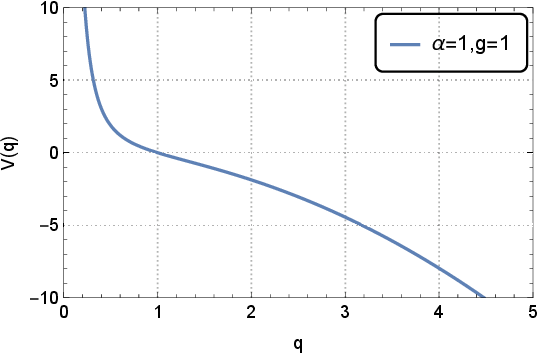}
  \caption{Graph $V_{\!_{S}}(q)$ vs $q$ of the classical effective potential associated with the operator $S$.
    The chosen parameters are
    $g=1$, $\alpha=1$. The potential is monotonic with no lower or upper bounds, and
      has two competing behaviors: an effective centrifugal-like barrier for small $q$ and an inverted harmonic oscillator behavior 
     as $q \rightarrow \infty$.}
     \label{fig:S-potential}
 \end{figure}

For the study of causal diamonds, we are especially interested in the physics generated by the rescaled 
hyperbolic operator $S$ and its dual operator $R$.
Specifically, the role of $S$ as time-evolution generator is discussed in the next section.
Now, from Eq.~(\ref{eq:tau-t-generic}), 
the associated effective time for $S$ is given by
\begin{equation}
    \tau_{\!_{S}} = \int \frac{dt}{1-t^2/\alpha^2} = \alpha \tanh^{-1} (t/\alpha)
    \; 
    \label{eq:tau-t-S}
\end{equation}
[with the initial condition  $\tau(t=0)=0$], and similarly $\tau_{\!_{R}} =  \alpha \tan^{-1} (t/\alpha) $ for $R$
(where $u= 1, v = 0, w = \mp\alpha^{-2}$, respectively).
 The corresponding transformed field variables 
${q}(\tau) \equiv q_{_{R,S}} (\tau)
= Q(t)/|1 \pm\left(t/\alpha \right)^2|^{1/2}
$ 
are defined in Eq.~(\ref{eq:q-Q-relation}).
Then, Eq.~(\ref{eq:Hamiltonians-for-G}) implies that the Hamiltonian representations
of the dual operators $R$ and $S$
 (with $\Delta = \mp 4/\alpha^2$)
  are 
\begin{equation}
         \left\{
    \begin{array}{c}
    R 
        \\
    S 
        \end{array}
    \right\}
      = \frac{1}{2} \, p_{\!_{R,S}}^{ 2} (\tau) + V_{\!_{R,S}}( q_{_{R,S}} )
    \; ,
    \label{eq:R-S-explicit} 
\end{equation}
 subject to the effective potentials
\begin{equation}
   V_{\!_{R,S}}( q ) = \frac{g}{2q^2} \pm \frac{1}{2\alpha^2}q^2
    \; ,
\label{eq:effective-potential_R-S}
\end{equation}
which can be identified as useful probes of their spectral properties
 (see also Refs.~\cite{dAFF,PI-CQM}). 
The effective potentials~(\ref{eq:effective-potential_R-S})
consist of a sum of the inverse square potential and a regular or inverted harmonic oscillator potential,
respectively. Thus, while the effective potential of the operator $R$ is a modified harmonic oscillator, that of the operator $S$ is unbounded, monotonic, and repulsive for all positive values of $q>0$; see
Fig.~\ref{fig:S-potential}.
As a result, the elliptic and hyperbolic families are distinctly different: 
\textit{elliptic operators are compact with a discrete spectrum, 
while hyperbolic operators are unbounded with a continuous spectrum\/}.

\section{Conformal Quantum Mechanics as the Time-Evolution Dynamics of Causal Diamonds}
\label{sec:causal-diamonds_CQM}

This section starts with a brief overview of the {\em hyperbolic operator $S$ as the generator of time
evolution within causal diamonds\/}. The discussion leads to a deeper analysis of the physics of diamond observers 
in terms of their dynamical time $\tau$, which pave the way for the next section addressing
the core development of this work: the central role played by the operator $S$
 in driving thermal properties of causal diamonds. 
 
\subsection{Time-Evolution dynamics of causal diamonds with radial conformal Killing fields}
\label{sec:causal-diamonds_time-evolution-RCKF}
 
The relationship between causal diamonds and the symmetry generators of CQM has been explored in 
Refs.~\cite{arzano,arzano2} 
in the context of radial conformal Killing fields (RCKF)~\cite{RCKF} in Minkowski spacetime.

The RCKFs are radial vector fields $\xi$ such that the Lie derivative $\mathcal{L}_{\xi}$ of the Minkowski metric
satisfies $\mathcal{L}_{\xi} \eta_{\mu \nu} \propto \eta_{\mu \nu}$.
 The most general RCKF is a linear combination of the following three relativistic Killing
generators in Minkowski spacetime, as shown in Ref.~\cite{RCKF}:
(i) the time-translation Killing vector $P_{0}=  \partial_{t}$,
(ii) the dilatation generator $D_{0} = r \partial_{r} +  t \partial_{t}$,
 and (iii) the special conformal transformation generator 
 $K_{0}= 2t r \partial_{r} +  (t^2 + r^2 ) \partial_{t}$ (with respect to the time direction).
As a cautionary remark, in this section, the time variable $t$ is the inertial Minkowski time, which is conceptually 
 different from (but analogous to) the CQM time $t$ of the other sections of this paper.
 Then, $\partial_t$ is the timelike Minkowski Killing vector; and also, $r>0$ is the radial distance from the origin.
 The Killing generators satisfy an $sl(2,\mathbb{R}) \approx so(2,1)$ algebra
 \begin{equation}
 [D_{0}, P_{0}] = - P_{0}
    \; , \qquad\qquad 
   [D_{0}, K_{0}] =  K_{0}
    \; , \qquad\qquad 
  [P_{0}, K_{0}] =  2 D_{0}
    \; ,
    \label{eq:sl2R-algebra_Killing}
 \end{equation}
 as a subset of the spacetime conformal algebra.
  The algebra~(\ref{eq:sl2R-algebra_Killing}) 
  (with the extra imaginary unit $i$ in the conversion from Killing fields to self-adjoint Lie generators),
 is in one-to-one correspondence with the CQM $sl(2,\mathbb{R})$ algebra of Eq.~(\ref{eq:sl2R-algebra}).
 Moreover, the statement that an RCKF is of the form 
 $\xi = a K_{0}+ b D_{0} + c P_{0}$, with linear-combination constants $a$, $b$, and $c$,
 is also in one-to-one correspondence with the CQM statement~(\ref{eq:generalized-generators}).
  Therefore, via this 
  \textit{CQM-RCKF operator correspondence}:
   \begin{quotation}
 \noindent
 CQM can be regarded as an effective theory 
 whose symmetry algebra and generator flows are isomorphic to those of the spacetime RCKFs.
 \end{quotation}
 From this isomorphism, one concludes~\cite{arzano,arzano2} that:
 (i) the flows of Killing vectors in Minkowski spacetime 
 can be faithfully represented with the corresponding Hamiltonian evolutions of CQM; 
 (ii) the CQM classification of (Hamiltonian) generators correspondingly applies to the RCKFs.
 These properties are further outlined in the next paragraph.
 
\begin{figure}[htb!]
    \centering
    \includegraphics[width=0.4\linewidth]{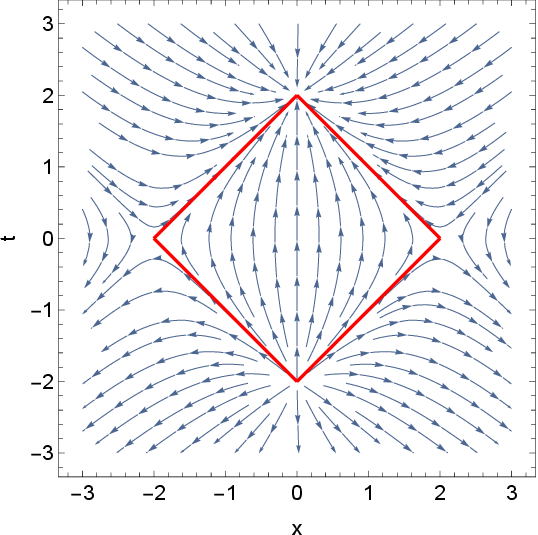}
    \caption{The integral curves of the RCKF operator $S_K$ (with $\alpha=2$) are shown. 
    The diamond-shaped region is the causal diamond.
     In this diagram, the variable
    $r$ is continued to both positive and negative values to cover the whole real line. 
    In particular, the  integral curves within the diamond stay within the diamond.}
    \label{fig:tr-streamline}
\end{figure}

 Given the CQM-RCKF operator correspondence, one can define the analogs of the $R$ and $S$ operators in the 
 form of Eq.~(\ref{eq:hat-R-S'-def}), supplemented by the spacetime dilatation generator $D_{0}$:
 \begin{equation}
 \begin{aligned}
 & R_{K}
  = \frac{1}{2}\left(\alpha P_{0} + \frac{1}{\alpha} \, K_{0} \right)
=  \frac{1}{2\alpha}\left[(\alpha^2+t^2+r^2)\partial_t +2tr\partial_r\right]
 \; ,
 \\
    & S_{K}  
    = \frac{1}{2}\left(\alpha P_{0} - \frac{1}{\alpha} \, K_{0} \right) 
    = \frac{1}{2\alpha}\left[(\alpha^2-t^2-r^2)\partial_t -2tr\partial_r\right]
    \; ,
 \\
  & D_K  
  \equiv D_{0} = r\partial_r + t\partial_t  
    \; .
 \label{eq:R_K-S_K-D_K}
  \end{aligned}
  \end{equation}  
 With the rescalings of Eqs.~(\ref{eq: rescaled-R})--(\ref{eq: rescaled-S}), and
 the notation $\approx$ for the CQM-RCKF operator correspondence,
 the correspondences are:
 $ 2R_{K}/\alpha  \approx  R/i 
 = \partial_{\tau_{_{R}}}$,
  $ 2S_{K}/\alpha  \approx  S/i 
  = \partial_{\tau_{_{S}}}$,
   and $D_{K} \approx D/i $. 
    In particular, the hyperbolic class of RCKFs, which has $S_{K} $ as a prototypical operator, is the one
 that maps the causal diamond of size $2\alpha$ into itself~\cite{DeLorenzo_light-cone,DeLorenzo_BH,arzano}.
 Therefore, the integral curves of the RCKF operators
  $S_K$ are timelike spacetime trajectories within the causal diamond (see Fig.~\ref{fig:tr-streamline});
     in terms of a timelike variable $\tau$, the operator 
  $S/i = \partial_\tau$ generates time-translations for a finite-lifetime observer 
  whose proper time is $\tau \equiv \tau_{_{S}}$. 
In a similar manner, the operators $R$ and $D$
 give elliptic and hyperbolic integral curves, respectively.
   
   As an added bonus, 
   Ref.~\cite{arzano} gives a direct connection between the two-point function of CQM~\cite{Jackiw_CQM-QFT}
   and static diamond observers based upon the SO(2,1) group-theoretic structure;
   this includes a derivation of the diamond temperature $T_D$ of Eq.~(\ref{eq:diamond-temperature}). 
However, despite the truly remarkable connection between causal diamonds and CQM of Refs.~\cite{arzano,arzano2},
  the physical meaning of the CQM two-point function remains elusive.
  
\subsection{Time-evolution dynamics of causal diamonds: The physics of diamond observers}
\label{sec:causal-diamonds_time-evolution-observers}

In this article, using the explicit Hamiltonian form of the $S$ operator 
 given in Eqs.~(\ref{eq:R-S-explicit})--(\ref{eq:effective-potential_R-S}),
 we find that the CQM symmetries and associated integral orbits
 lead to the diamond temperature  $T_D$ of Eq.~(\ref{eq:diamond-temperature}).
 Most importantly, we also develop  a fully quantum-mechanical
description of the underlying physics. These novel results will be shown in the next section.
As a first step, while the RCKF operator $S_K$ of Eq.~(\ref{eq:R_K-S_K-D_K}) provides a direct link 
with causal diamonds from its integral curves, the physics of a finite-lifetime observer
 does not appear to have an obvious connection with its representation in 
 Eqs.~(\ref{eq:R-S-explicit})--(\ref{eq:effective-potential_R-S}).
 This connection can be understood in terms of the relationship between the time-parameters 
 $\tau \equiv \tau_{S}$ and $t$ (associated with $S$ and $H$ respectively). 
 In this relationship, Eq.~(\ref{eq:tau-t-S}), the interval 
 $\tau\in(-\infty,\infty)$ is mapped into the finite interval $t\in[-\alpha,\alpha]$. 
Correspondingly, if $H$ is the Hamiltonian of a system in Minkowski spacetime, the action of the hyperbolic
 generator $S$ will create a trajectory restricted to the Minkowski time $t\in[-\alpha,\alpha]$. 
 Thus, for a system whose time evolution is generated by $S$, 
causal access is restricted within the causal diamond shown in Fig.~\ref{fig:causal-diamond}.
 As a consequence, Eq.~(\ref{eq:tau-t-S}) defines the time parameter for the trajectory of a 
 finite-lifetime observer inside a causal diamond, 
 which is the so-called diamond time~\cite{Su2016SpacetimeDiamonds,camblong-ent-deg-2024},
 $t$ is the ``inertial'' Minkowski time, and $2\alpha$ is the size of the diamond.  
 Moreover, these times coincide ($\tau = t$) in the limit $\alpha \rightarrow \infty$, i.e., when the observer's 
 lifetime becomes infinitely large, approaching the physics of ordinary Minkowski spacetime.
 This restatement of the properties of the hyperbolic operator $S$ shows that it can be interpreted rigorously
  as the generator of time evolution for a diamond observer.

\section{Quantum analysis of thermality via path integrals}
\label{sec:semiclassical_GF-thermality}

This section presents the most important result of our paper: 
{\em the thermal nature of causal diamonds, driven by the properties of the 
hyperbolic operator $S$ as the generator of its time evolution.\/}
Our complete analysis involves path integral and Green's function techniques. In addition to addressing 
the relevant quantum features of the thermality of causal diamonds, we will examine some ingredients of its semiclassical regime, which has the seeds of the full-fledged quantum behavior.

It is noteworthy that the operator $S$ is noncompact, with a spectrum unbounded from below. As a result,
it exhibits a dynamical instability, which can be formally discussed both in terms
of a semiclassical method and using an out-of-time-order correlation (OTOC) function~\cite{OTOC3,OTOC5};
with these methods, a growth scrambling rate (quantum Lyapunov exponent)
 $\lambda_L$ can be extracted, as elaborated upon in 
 Sec.~\ref{sec:semiclassical_instability-thermality}.
  Moreover, as in the case of black holes, this appears to be related to their thermal nature~\cite{majhi-nh-thermality1}.
  
  In this section, and for the remainder of the paper, we restore the reduced 
Planck's constant $\hbar$ in all the relevant equations.

  \subsection{Hyperbolic-elliptic operator duality via analytic-continuation}
\label{subsec:analytic-continuation}

For our purposes, despite the central role played by the CQM hyperbolic generator $S$ in our discussion, 
a critical step of the thermality analysis will be carried out with the elliptic operator $R$, which has some advantages related to its explicit avoidance of the instability behavior of the operator $S$.
 The technical use of one operator in lieu of the other is justified by the form of 
  the prototypical elliptic and hyperbolic operators $R$ and $S$, as written
 in Eqs.~(\ref{eq: rescaled-R})-(\ref{eq: rescaled-S}); these equations
show that the following analytic extensions with respect to $\alpha$
 map $R$ into $S$, and vice versa~\cite{PI-CQM}:
    \begin{equation}
   \begin{aligned}
   &
    S = R \left[ \omega_{\!_{R}} \equiv  \alpha_{\!_{R}}^{-1}
    \longrightarrow  \omega_{\!_{R}} \equiv  \alpha_{\!_{R}}^{-1} = -i \omega_{\!_{S}} \equiv - i \alpha_{\!_{S}}^{-1} \right]
       \\
   &
       R = S \left[ \omega_{\!_{S}} \equiv  \alpha_{\!_{S}}^{-1}
    \longrightarrow  \omega_{\!_{S}} \equiv  \alpha_{\!_{S}}^{-1} = i \omega_{\!_{R}} \equiv i \alpha_{\!_{R}}^{-1} \right]
      \end{aligned}
\; .
  \label{eq:S-to-R_alpha}
    \end{equation}
The frequency parameter $\omega$ will be used to interpret the associated effective potential behavior.
 Equation~(\ref{eq:S-to-R_alpha}) involves the specific analytic continuation that has all the desirable properties,  
 including the critical extrapolation of the asymptotic behaviors and Green's functions from one sector of the theory to another~\cite{PI-CQM}.
 Furthermore, Eq.~(\ref{eq:S-to-R_alpha}) defines an 
 \textit{operator duality transformation between the generators $R$ and $S$\/}, and, more generally, 
 between the hyperbolic and elliptic families.
 
One of the technical advantages of the operator $R$ is the fact that it admits 
closed periodic orbits in phase space due to its compact nature. 
This feature can be most easily analyzed 
with the Hamiltonian representations of Eqs.~(\ref{eq:R-S-explicit})--(\ref{eq:effective-potential_R-S}),
which can be rewritten as
\begin{equation}
    \left\{
    \begin{array}{c}
    R 
        \\
    S 
        \end{array}
    \right\}
     =
     \frac{1}{2}\, p_{\!_{R,S}}^2      + \frac{g}{2q_{\!_{R,S}}^2 } 
     \pm \frac{1}{2{\alpha}^2} \, q_{\!_{R,S}}^2  
     \; .
     \label{eq:Hamiltonians-for-R-S}
     \end{equation}
The effective times $ \tau_{\!_{R,S}}$ as well as the transformed field variables 
$ q_{\!_{R,S}}$ and conjugate momenta $ p_{\!_{R,S}}$ are different for $R$ and $S$, according to 
 Eqs.~(\ref{eq:tau-t-generic})--(\ref{eq:q-Q-relation}), with their explicit forms in and around Eq.~(\ref{eq:tau-t-S}).
 The parameter $\alpha$ can be interpreted as an effective inverse frequency $\omega$,
i.e., $| \omega |= 1/\alpha$. 
Despite the apparent complexity,
the analytic continuation of Eq.~(\ref{eq:S-to-R_alpha})
adjusts all the variables to their respective natural values.
Most importantly, the transformed effective times, field variables,
and their associated effective energies enter the ensuing path-integral functionals 
automatically.

\subsection{Path-integral amplitude functionals: Canonical and microcanonical}
\label{subsec:PI-functionals}

Another important set of tools in our analysis is the use of path-integral expressions for the probability 
amplitudes~\cite{Kleinert-PI,Grosche-PI}. 
These are defined in terms of exponentials of action functionals summed over configuration paths. 
In particular, there are two important action functionals at the classical level: the ordinary 
Lagrangian action associated with Hamilton's principle and the Jabobi action associated with the Maupertuis principle.
These lead to two distinct but related path-integral expressions at the quantum level. 
In addition, the path-integral results can be derived either in full-fledged quantum mechanical form or via
 asymptotic semiclassical approximations.
In what follows, we will use appropriate ingredients of these miscellaneous approaches in the arguments supporting the thermality of causal diamonds.

Hamilton's formulation of the action functional $S[q(t)] = \int L(q,\dot{q},t) dt$,
associated with the variational Hamilton's principle,
gives a value of $S[q(t)]$ that depends on fixed configuration end points
$q'$ and $q''$ with given endpoint times $t'$ and $t''$;
in particular, this value only depends on the time difference $T=t''-t'$ for conservative systems.  
This feature can be emphasized by writing the action functional (or Hamilton's principal function) 
as $S[q(t)](q'',q';T)$, for which we will use the shorthand $S[q;T]$,
 highlighting that $T$ is fixed but $E$ is not. 
 (In this section, we will consider $q$ to be a one-dimensional coordinate, though it could represent 
 a system with an arbitrary number of coordinates, and this is relevant for CQM; 
 see Appendix~\ref{sec:app_PI-functionals}).
 Moreover, from the variational conditions satisfied in classical dynamics~\cite{Landau_ClassMech,Goldstein_ClassMech,Lanczos_VariationalMech}, the following time derivative gives one of the fundamental equations:
 $-\partial S[q;T]/\partial T =  E$; this statement is applicable to the Hamiltonian taking values $H=E$.
 
Alternatively, Jacobi's action formulation for conservative systems~\cite{Landau_ClassMech,Goldstein_ClassMech,Lanczos_VariationalMech}
involves the combination 
\begin{equation}
W[q(t)](q'',q';E) \equiv
W[q;E]= S[q;T]+ET
\label{eq:Jacobi-action-as-Legendre}
\; ,
\end{equation}
which is a Legendre transform that exchanges the variables $E$ and $T$.
Thus, this functional $W[q;E]$, called the Jacobi action,
 Hamilton's characteristic function, or abbreviated action, 
 has a fixed energy $E$ for all paths, but the corresponding physical time interval $T$ between initial and final configurations is path dependent~\cite{Landau_ClassMech,Goldstein_ClassMech,Lanczos_VariationalMech}.
The Jacobi action so defined can be rewritten, 
from the definition of the Hamiltonian, 
as the momentum 
integral with respect to the configuration-space path, 
\begin{equation}
   W[q;E]
   =
    \int  p dq
    \; ,
    \label{eq:Jacobi-action-explicit}
\end{equation}
 and it is associated with the variational Maupertuis principle.
Even though the Jacobi formulation of the action principle does not involve the time explicitly, the {\em conjugate
classical time interval\/} can be determined from the ensuing classical equations of motion,
 and is given from the Legendre transform by 
\begin{equation}
     T = \frac{\partial W[q;E]}{\partial E}
    \; .
    \label{eq:time-from-Jacobi-action}
\end{equation}
This formulation
  has important applications in gravitational theory vis-a-vis the general-relativistic definition of time~\cite{Hartle_GRtime,Brown-York_GRtime,Brown-York_microcanical,Brown-York_DOS},
and is central to semiclassical descriptions of quantum mechanics and path integrals, as in the context of quantum chaos~\cite{Gutzwiller_GTF,Gutzwiller_Chaos-ClassQM,Haake_QChaos,Stockmann_QChaos}.

For the quantum-theory path integrals, these two frameworks (Hamilton's versus Jacobi's action) correspond to the configuration-space representations of the transition amplitude $K(q'',q';T)$ and 
 the energy Green's function $G(q'',q';E)$ respectively.
The transition amplitude is given by the standard path integral
\begin{equation}
K(q'',q';T)   
= 
\int_{  {q} (t')  = {q}'  }^{  {q} (t'')  = {q}'' }
 \;  
{\cal D} {q} (t) \,
\exp \left\{ 
\frac{i}{\hbar} 
S \left[ {q}(t)  \right]  ({q}'', {q}' ; T)  
\right\}
\; ;
    \label{eq:propagators-def_main-text}
\end{equation}
and the (retarded) energy Green's function is defined 
by
\begin{equation}
G (q'',q';E) 
    = 
     \frac{1}{i \hbar} \int_{0}^{\infty} d T \, 
    \int_{  {q} (t')  = {q}'  }^{  {q} (t'')  = {q}'' }
 \;  
{\cal D} {q} (t) \,
\exp \left\{ 
\frac{i}{\hbar} 
W \left[ {q}(t)  \right]  ({q}'', {q}' ; E)  
\right\}
    \; .
    \label{eq:Green-functions-def_main-text}
\end{equation}
It should be noted that the energy Green's functions involve a path integral with a measure that includes an extra $T$ integration, and unlike 
the ordinary path integral of Eq.~(\ref{eq:propagators-def_main-text}), 
they are effectively computed with the Jacobi action---this can be further developed using the theory of constraints to characterize all the energy functionals in a compact 
formulation~\cite{Brown-York_GRtime,Brown-York_microcanical,Brown-York_DOS}.
In the thermal interpretation of quantum physics with imaginary time, these two frameworks are the 
{\em canonical and microcanonical actions and path integrals.\/}
We will use these frameworks below to derive the basic thermal properties of causal diamonds, 
with supporting details and generalizations in Appendix~\ref{sec:app_PI-functionals}.

In addition, by considering periodic boundary conditions where $q (t'') = q(t')$, one can define formal 
traces of the above operators, which have been called quantum-mechanical 
partition functions~\cite{Kleinert-PI}---and
indeed, they correspond to the statistical-mechanical partition functions when the analytic extension 
to imaginary time defines thermal functionals. This is used in the discussion of thermal properties below. 
The canonical and microcanonical trace functions are defined 
as the operator traces (associated with a Hamiltonian $\hat{H}$)
\begin{align}
&
\tilde{Z}^{(H)} (T)  
\equiv {\rm Tr} [K^{(H)}(T)]
 \equiv {\rm Tr} \left[  e^{-i \hat{H} T/\hbar} \right] 
= \int d q' K^{(H)}(q',q'; T) 
\; ,
\label{eq:trace-K}
\\
&
{\rm Tr} \left[ \hat{G}^{(H)}(E)  \right]
=
\frac{1}{i \hbar} \int_{0}^{\infty} d T 
  \, e^{iET/\hbar}\; \tilde{Z}^{(H)} (T)  
= \int d q' G^{(H)}(q',q'; E) 
\label{eq:trace-G}
\; .
\end{align}
We can refer to these quantities~(\ref{eq:trace-K}) and (\ref{eq:trace-G})
as the ``traces of the corresponding amplitude functions,'' i.e., 
${\rm Tr} [K(T)] $ and ${\rm Tr } [G(E)]$.
As a byproduct of these definitions, 
one can also define various derived statistical quantities, including
the density of states (see Appendix~\ref{sec:app_CQM-generators-summary}).

\subsection{Thermality of causal diamonds
 from elliptic operator \boldmath{$R$}: Duality, period and analytic continuation}
\label{subsec:thermality-from-duality}

We now refer back to the analytic continuation~(\ref{eq:S-to-R_alpha}) 
defining a duality between the operators $S$ and $R$.
The goal is to take advantage of the property that the operator R admits closed periodic orbits in classical phase space,
which we will label with a subscript $\gamma$.
These trajectories can be used generally in the quantum-mechanical traces~(\ref{eq:trace-K}) and (\ref{eq:trace-G})
and related statistical properties (as shown below), and they play an important role for both classical and quantum chaos.
However, the operator $S$ is the relevant one for the 
 causal-diamond dynamics (as shown in Sec.~\ref{sec:causal-diamonds_CQM}).
Thus, our procedure consists of the following steps:
\begin{enumerate}
\item[(i)] start with the operator $S$, and 
analytically continue 
it to $R$ with the duality of Eq.~(\ref{eq:S-to-R_alpha});
\item[(ii)]
compute the required quantities 
(e.g., Jacobi action and classical time period) for the phase-space
periodic orbits $\gamma$ corresponding to $R$ as Hamiltonian function;
\item[(iii)]
analytically continue the expressions for the operator $R$ back into the
 corresponding expressions for $S$.
 \end{enumerate}
 In order to implement this scheme, we will make the analytic continuations of Eq.~(\ref{eq:S-to-R_alpha}) 
 more concrete, with a {\em specified parametrization\/} that identifies the correct $\alpha$ in each sector of the theory.
 Then, we can rewrite definitions of
 $R$ and $S$ in Eq.~(\ref{eq:Hamiltonians-for-R-S}) considering a specific analytic continuation,
 \begin{equation}
 \begin{aligned}
  &
  S=     \frac{1}{2} \, p^2 + \frac{g}{2q^2} - \frac{1}{2 \alpha_{\!_{S}}^2 }q^2
  \; \;  \; \; (\text{where} \; g,\alpha_{\!_{S}} \in \mathbb{R})
  \\
&    R=  S \left[ \alpha_{\!_{S}}^{-1} = i \alpha_{\!_{R}}^{-1} \right]
=
   \frac{1}{2}\, p^2 + \frac{g}{2q^2} +  \frac{1}{2  \alpha_{\!_{R}}^2}q^2
         \; ,
     \label{eq:Hamiltonians-for-R-S_redef}
\end{aligned}
\end{equation}
with the understanding that this transforms the field variables from 
$q_{\!_{S}}$ to $q_{\!_{R}}$, and momenta from $p_{\!_{S}}$ to $p_{\!_{R}}$, along
with the corresponding effective times ($\tau_{\!_{S}}$ and $\tau_{\!_{R}}$).
This is a useful transitional notation for the
interpretation of the results of the thermal nature of causal diamonds. As shown below,
this parametrization will be used temporarily until
the final appropriate analytic continuation is enforced. 
 This procedure takes care of step (i). Step (ii) is implemented below 
 with periodic orbits of $R$ as if $\alpha_{\!_{R}}$ were real;
 and in step (iii) an appropriate replacement
leads to the final identification of a real, geometrical $\alpha$ 
parameter (i.e, $\alpha \in \mathbb{R})$.

In particular, for step (ii), from Eq.~(\ref{eq:Hamiltonians-for-R-S}),
the classical trajectories of the Hamiltonian operator $R= R(q,p)$ 
with energy $E_{\gamma}$ follow the equation 
\begin{equation}
R(q,p)
=
     \frac{1}{2}\, p^2 + \frac{g}{2q^2} +  \frac{1}{2  \alpha_{\!_{R}}^2}q^2
      = E_{\gamma}
    \; .
\label{eq:Hamiltonian-orbits-for-R}
\end{equation}
It should noted that this construction, based on periodic orbits, is not feasible with the original operator $S$, whose space trajectories are unbounded. (A lucid description of this for the similar case of the simple inverted harmonic oscillator can be found in Ref.~\cite{log-phase-sing_IHO}.)
The time period $T_{\gamma}$ of these closed orbits can then be determined either from the general equation
for the conjugate classical transit time, Eq.~(\ref{eq:time-from-Jacobi-action}), after computing the Jacobi action
$W(E_{\gamma})$, as shown in Sec.~\ref{sec:semiclassical_instability-thermality};
or from a direct application of the definition of transit time~\cite{Kleinert-PI}
\begin{equation}
  T(E) = 2 
  \int_{q_{-}}^{q_{+}} \,  
 \frac{dq}{ \displaystyle \bigl| p(q;E) \bigr|   }
 \; ,
 \label{eq:1D-period}
 \end{equation}
 (with effective mass $M=1$ in the notation of Appendix~\ref{sec:app_CQM-generators-summary}),
where $p(q;E)$ is defined via Eq.~(\ref{eq:Hamiltonian-orbits-for-R}), 
and the integral is between the turning points $q_{\mp}$;
see Appendix~\ref{sec:app_DOS_operator-S}, Eq.~(\ref{eq:1D-period-DOS}), for additional details. 
The integral in Eq.~(\ref{eq:1D-period}) is straightforward, and the resulting period is
\begin{equation}
    T_{\gamma} =  \pi \alpha_{\!_{R}}
    \; 
    \label{eq:period-from-action}
    \; ,
\end{equation}
which is proportional to the oscillator parameter $\alpha_{\!_{R}}$.

So far, the results above for periodic orbits are specific to the operator $R$. However,
 our ultimate goal is to find the corresponding statements for the hyperbolic operator $S$. 
It is at this critical step (iii) that an appropriate interpretation is needed for every chosen framework, either
  microcanonical or canonical.
 Here, we can formulate the microcanonical result, with the canonical one discussed in the next section.
 The microcanonical statement follows by analytic continuation back to the original
  $\alpha_{\!_{S}} $, which is the 
 real parameter $\alpha$,
thus implying that
 $\alpha_{\!_{R}} = i\alpha_{\!_{S}} = i \alpha$.
 Therefore, as a result of the analytic extension relating the hyperbolic and elliptic sectors of the theory,
 the time period $T_{\gamma}$ now has acquired an imaginary value
\begin{equation}
     \left.
     T_{\gamma}^{\!\,}
      \right|_{{\text{microcanonical}}}
     =
\left.
T_{\!_{R}}^{\!\,}
 \right|_{{\text{microcanonical}}}
     = i\pi\alpha
    \; .
\label{eq:R-period-microcanonical}
\end{equation}
The statement about the period in Eq.~(\ref{eq:R-period-microcanonical}) can be made for any 
interpretation of the theory in the microcanonical ensemble; for example,
it can be used for the semiclassical density of states 
$\rho(E)$, as shown in Appendix~\ref{sec:app_DOS_operator-S}.
The fact that Eq.~(\ref{eq:R-period-microcanonical}) is an imaginary time 
points to a thermal interpretation of the state of the system.
However, the correct value of the temperature has to be extracted from the operator $S$ directly, and 
a valid procedure involves an equilibrium state at a given temperature, i.e., the canonical ensemble, 
as is discussed next.

\subsection{Thermality of causal diamonds: Canonical-ensemble characterization 
and conclusions}
\label{subsec:thermality-conclusions}

The physical meaning of the imaginary time period, which was found in the previous section,
 is correctly given by the formal correspondence between time and inverse temperature within
  the general framework for the statistical mechanics of a thermal system.
 In this framework, the Wick rotation to Euclidean time is defined in the {\em canonical ensemble\/},
and leads to an identification of the imaginary time with a parameter $\beta$ as the inverse temperature~\cite{Wick-rotation}. 

The expression for the period 
in Eq.~(\ref{eq:R-period-microcanonical}) is only a particular case for classical periodic orbits of $R$.
By contrast, in the canonical ensemble, one has a more general time interval $T \equiv T_{_{S}}$ in the dynamics of the operator $S$.
Then,
the {\em correct and fully general statement for the time interval $T$ is made in terms of the analytic continuation of
 the path-integral transition amplitude for the causal-diamond evolution operator $S$\/}
 (our object of interest).
As shown in Appendix~\ref{sec:app_CQM-generators-summary},
Eqs.~(\ref{eq:propagator-trace-relation_R-to-S})--(\ref{eq:propagator-trace-relation_explicit}),
the traces
$\tilde{Z}^{(S)}(T)$ and $\tilde{Z}^{(R_{\rm Eucl})}(T)$ are identical:
\begin{equation}
\tilde{Z}^{(S)}(T)
= \tilde{Z}^{(R_{\rm Eucl})}(T)
\; .
\label{eq:partition-function-S_01}
\end{equation}
Specifically, these identical functions
permit an extension of the results for the operator $R$ into meaningful results for the operator $S$. 
 To be more precise, the Euclidean-time replacement can be written as 
$ T_{_{R}}
= - i  T_{_{R_{\rm Eucl}}}$, 
and this implies $ T_{_{S}} =T_{_{R_{\rm Eucl}}}$ in Eq.~(\ref{eq:partition-function-S_01}). 
In addition, the thermal assignment of the Euclideanized theory for $R$ 
implies that the use of the time
in $\tilde{Z}^{(R_{\rm Eucl})}(T)$ conforms to the
 replacement $T \rightarrow -iT$, 
 where the imaginary time is identified as an inverse temperature $\beta$,
 i.e.,
 \begin{equation}
 T_{_{S}} =
 T_{_{R_{\rm Eucl}}} = \hbar \beta
 \; .
 \label{eq:time-temperature-identification}
 \end{equation}
Thus, from Eqs.~(\ref{eq:time-temperature-identification}) and 
(\ref{eq:propagator-trace-relation_R-to-S})--(\ref{eq:propagator-trace-relation_explicit}),
\begin{equation}
{Z}^{(S)}( \beta) \equiv
\tilde{Z}^{(S)}(\hbar \beta)
= \tilde{Z}^{(R_{\rm Eucl})}(\hbar \beta)
=
\frac{e^{-\mu\beta \hbar \omega }}{2\sinh \left( \beta \hbar \omega \right)}
\; ,
\label{eq:partition-function-S_01-b}
\end{equation}
which can be interpreted as giving an ordinary statistical partition function ${Z}^{(S)}(\beta)$ 
 in the canonical ensemble for the operator $S$ as the physical time-evolution Hamiltonian. 
 In Eq.~(\ref{eq:partition-function-S_01}), as in Appendix~\ref{sec:app_CQM-generators-summary}, 
 Eq.~(\ref{eq:1D-R-propagator}),
 the conformal index is $\mu= \sqrt{g+1/4}$ for the 1D dAFF model.
 
 It should be noted again that this logical argument uses an analytic continuation of
 the propagator of the operator $S$, which reduces to the computation of relevant quantities of the corresponding 
 amplitude for the compact operator $R$.
 Then, in Eq.~(\ref{eq:partition-function-S_01-b}),
 the denominator $2\sinh \xi 
 = e^{\xi} \left( 1- e^{-2\xi} \right) $
 can be expanded in a geometric series, 
whence, 
\begin{equation}
{Z}^{(S)}(\beta)
=
\sum_{n=0}^{\infty}
e^{- \beta \hbar \omega   \left(2n + \mu + 1 \right)}
=
\sum_{n=0}^{\infty}
e^{-\beta  E_{n} }
\; ,
\label{eq:partition-function-S_02}
\end{equation}
where the replacement
 \begin{equation}
 E_{n} = \hbar \omega \left(2n + \mu + 1 \right) 
 \label{eq:partition-function_eigenvalues}
\end{equation}
 is made.  
 Equation~(\ref{eq:partition-function_eigenvalues})
 gives the known eigenvalues of the operator $R$ (equivalent to 
 a combination of an ordinary harmonic oscillator with an inverse square potential)~\cite{PI-CQM}.
Then, this shows that the partition function or trace of the propagator indeed corresponds to the canonical-ensemble
expansion and a {\em thermal quantum density operator\/}
\begin{equation}
\hat{\rho} (\beta)
=
e^{-\beta R }
\; ,
\label{eq:partition-function-S_03}
\end{equation}
such that ${Z}^{(S)}(\beta) = {\rm Tr}  \left[ \hat{\rho} (\beta) \right] $,
thus providing a {\em characterization of the thermal nature of causal diamonds.\/} 

Even though Eqs.~(\ref{eq:partition-function-S_01})--(\ref{eq:partition-function-S_03})
realize a thermal interpretation, the value of the temperature is still unspecified. 
At this level, other arguments, already known in the literature, could be 
used~\cite{martinetti-1,Su2016SpacetimeDiamonds,Chakraborty:2022oqs,Foo2020GeneratingMirror,arzano,arzano2,camblong-ent-deg-2024,Tada_CQM-SSD} to determine the temperature.
However, this determination is also contained in the current duality framework, with the results of the previous section,
in the limiting semiclassical regime.
Indeed, the semiclassical periodic orbits in phase space 
are associated with a particular case of the general quantum-mechanical paths in the functional integrals
 leading to the trace operation that defines the partition function 
$\tilde{Z}^{(R_{\rm Eucl})}(\beta)$ in the Euclideanized version of operator $R$.
Therefore, the time in Eq.~(\ref{eq:period-from-action}) should be interpreted in the canonical ensemble
as giving a periodic path integral corresponding to the partition function, in such a way that 
\begin{equation}
   \left. T_{\!_{R_{\rm Eucl}}} \! 
   \right|_{{\text{canonical}}}
     = \pi\alpha
    \; ;
\label{eq:R-period-canonical}
\end{equation}
thus, Eqs.~(\ref{eq:time-temperature-identification}) and (\ref{eq:R-period-canonical}) yield
\begin{equation}
    \beta^{-1} = \frac{\hbar}{\pi \alpha} \equiv T_{D}
    \; .
    \label{eq:diamond-temp_derivation}
\end{equation}
This is exactly the anticipated temperature of a causal diamond $T_D$ with length $2\alpha$, 
Eq.~(\ref{eq:diamond-temperature}). 
(Cautionary note on notation: we are simply using $T$ to refer to a time interval, not a temperature. 
Instead, temperatures will be denoted via the inverse-temperature parameter $\beta$, 
with the only qualified exception of the diamond's temperature $T_{D}$.)

In short, our
derivation of thermality has been centered on the CQM effective-time-evolution operator $S$
and its analytic continuation to the operator $R$.
Most importantly, the analysis leading to Eq.~(\ref{eq:diamond-temp_derivation})
 shows both the existence of a {\em direct connection of CQM with the physics of 
finite-lifetime observers
and the remarkable thermal nature of causal diamonds.\/}

A remark about the generality of the proof of thermality follows.
The identification of the temperature $T_{D}$ is more general
 than the simple semiclassical analysis via Eq.~(\ref{eq:period-from-action}) would suggest.
 In effect, the analytic continuation~(\ref{eq:Hamiltonians-for-R-S_redef})
 suffices for the existence of a thermal character because
 it defines an {\em extension of the time for the operator $R$ into a thermal inverse temperature $\beta$.
This can be regarded as the primary condition of thermality.\/}
Effectively, the operator $R$ provides a more regular spectrum without instabilities, as in 
Eq.~(\ref{eq:partition-function_eigenvalues}),
 from which the partition function~(\ref{eq:partition-function-S_02}) 
 can be directly computed and the thermal density matrix~(\ref{eq:partition-function-S_03}) can be identified.
 Once the thermal character is established, the remaining task is to determine the
 otherwise unknown value of the temperature, which is 
 derived in terms of the causal-diamond size via the classical formula~(\ref{eq:period-from-action}) 
 for the period.
In this context, the {\em role played by the semiclassical analysis is simply to extract 
the unique value of the temperature, whose existence is already guaranteed
by the general framework.\/}

In conclusion, the thermal nature of the causal diamond and the
diamond temperature, Eq.~(\ref{eq:diamond-temperature}), as shown in this section,
are in agreement with results known using alternative approaches.
These findings correspond to an identification of causal 
diamonds with the thermal density matrix of Eq.~(\ref{eq:partition-function-S_03});
similar results were found in Ref.~\cite{Tada_CQM-SSD} using modular theory, without reference to causal diamonds, 
for the generic dAFF model. 
A complete determination of the thermality of causal diamonds with temperature~(\ref{eq:diamond-temperature})
 was first established in Ref.~\cite{martinetti-1},
 through the original Kubo-Martin-Schwinger (KMS) condition~\cite{KMS_Kubo,KMS_MS}
  in modular theory~\cite{Bisognano-Wichmann, Hislop-Longo,Haag_LQP};
 and it was confirmed with open quantum systems~\cite{Chakraborty:2022oqs},
Bogoliubov coefficients~\cite{Su2016SpacetimeDiamonds,Foo2020GeneratingMirror,camblong-ent-deg-2024}, and
group-theoretical methods~\cite{arzano,arzano2}.

\section{Semiclassical Aspects of Instability and Thermality in Causal Diamonds}
\label{sec:semiclassical_instability-thermality}

In this section, we develop another key result of our paper:
{\em the phase-space dynamics of the hyperbolic operator $S$ shows an instability behavior\/}
that appears to be related to the thermal properties of causal diamonds.

We obtain this behavior by first studying the 
 $S$-driven Hamiltonian dynamics, leading to the determination of the Lyapunov exponent. 
In addition, we fully characterize the closed phase-space orbits and associated Jacobi action, rederiving some 
of the thermality results of the previous section.
 This analysis concludes with basic results on information scrambling.

Additional semiclassical aspects of causal diamonds are explored  
in Appendix~\ref{sec:app_DOS_operator-S} via the density of states of the operator $S$, 
along with the corresponding fully quantum-mechanical results.

\subsection{Hamiltonian dynamics of the \boldmath{$S$} operator}
\label{subsec:S-Hamiltonian-dynamics}

The phase-space analysis is based on the identification of the operator $S$ as the Hamiltonian for
the time evolution governed by the natural time parameter
$  \tau_{\!_{S}}$,
 as was formulated in the previous section.
The system's Hamiltonian dynamics can then be fully characterized using Hamilton's equation of motion,
\begin{equation}
    \dot{q} = \frac{\partial S}{\partial p}  \;,\qquad\qquad \dot{p} = -\frac{\partial S}{\partial q} \; ,
\end{equation}
where $p $ is the canonical conjugate momentum. 
Therefore, with the explicit form of the operator $S$ in Eq.~(\ref{eq:Hamiltonians-for-R-S}), Hamilton's equations read
\begin{equation}
    \dot{q} = p\;,\qquad\qquad \dot{p} = \frac{g}{q^3} + \frac{q}{\alpha^2}
    \; .
    \label{eq:eq-of-motion-S}
\end{equation}
These equations are straightforward, but the behavior of the term ${q}/\alpha^2$ in the second one has
dramatic consequences for the stability analysis of the theory, as we show next.

\subsection{Phase-space analysis of the \boldmath{$S$} operator: Lyapunov exponent}
\label{subsec:phase-space_Lyapunov}

\begin{figure}[htb!]
    \centering
    \includegraphics[width=0.5\linewidth]{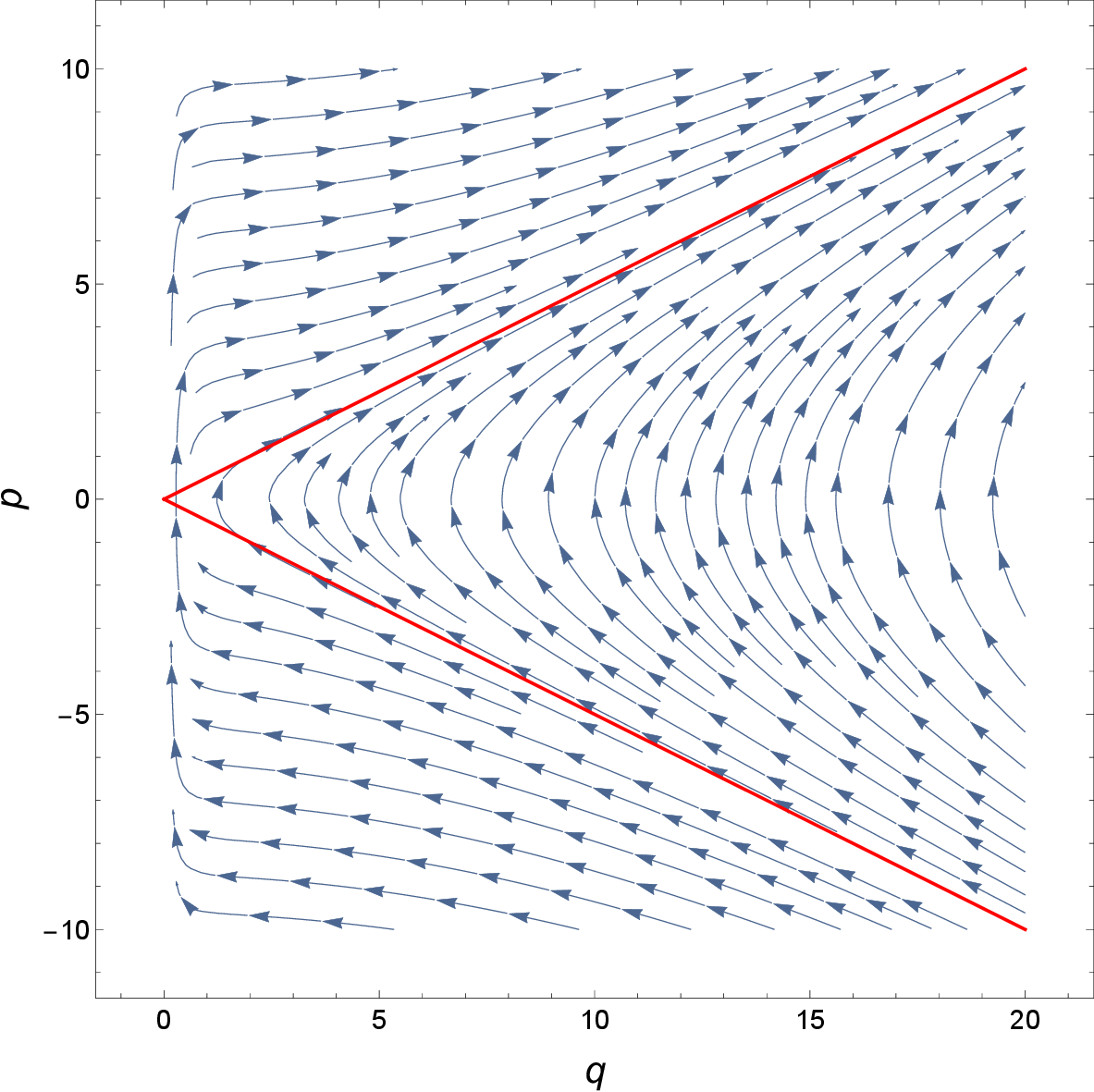}
    \caption{Phase-space direction field for the trajectories for the $S$ operator (with $g=1, \alpha =2$). The red line $p=q/\alpha$ shows the asymptotes of the trajectories.}
    \label{fig:phase-space-S}
\end{figure}

The phase-space dynamics of the hyperbolic operator $S$
can be understood by its representation with phase-space trajectories.
 These direction-field orbits are shown in Fig.~\ref{fig:phase-space-S}, with the asymptotes drawn in red. 

An important property for our purposes is the nature of the stability of the phase-space orbits.
This question can be answered by analyzing the evolution of a perturbation $\delta q(\tau) $
away from a given orbit $q(\tau) $.
In particular, if the perturbation grows exponentially, then the orbit is unstable, with a 
Lyapunov exponent or instability factor 
\begin{equation}
    \lambda_L = \lim_{\tau\rightarrow\infty} \frac{1}{\tau} \ln\left[\frac{\delta q(\tau)}{\delta q(0)}\right] 
    \; ,
\end{equation} 
which is completely determined by the long-term behavior of the perturbation,
 $\delta(q)(\tau\rightarrow\infty)$. 
 This asymptotic behavior can be determined via an effective-potential 
 analysis, considering 
 that $S$ has a purely repulsive behavior, with an infinite 
 barrier ($\propto 1/q^2$) near the origin (for $q \approx 0$)
  and an infinite well ($\propto - q^2$) near infinity 
   (for $q \approx \infty$);
   thus, we expect that $q$ will monotonically approach  $q \approx \infty$
   for sufficiently large values of $\tau$.
   This asymptotic behavior sets in when $\tau\gg \alpha$ and $q^4 \gg g \alpha^2$, 
and reduces Eq.~(\ref{eq:eq-of-motion-S}) self-consistently to
\begin{equation}
    \dot{q} = p \;,\qquad\qquad \dot{p} \approx  \frac{q}{\alpha^2}
    \; ,
    \label{eq:delta-q_growth}
\end{equation}
which leads to the late-time perturbation 
\begin{equation}
   \bigl.  \delta q(
    \tau\gg \alpha 
    ) \bigr|_{q^4 \gg g \alpha^2}
     \approx \delta q(0) \,e^{\tau/\alpha}
    \; .
\end{equation}
As a result, the Lyapunov exponent is given by
\begin{equation}
    \lambda_L = \frac{1}{\alpha} > 0
    \; .
    \label{eq:Lyapunov_causal-diamonds}
\end{equation}
In conclusion, the positive instability factor of Eq.~(\ref{eq:Lyapunov_causal-diamonds})
characterizes a perturbation that grows exponentially in the $q$ direction, due to a repulsive interaction, as in 
Eq.~(\ref{eq:delta-q_growth}), for any late-time orbit.
This shows that the system is unstable, supporting the expected intuition for a potential dominated at late
times by an inverted oscillator.
This dynamical instability is closely related to the thermal behavior of causal diamonds, as we will see next.

\subsection{Semiclassical analysis: Jacobi action of \boldmath{$R$}}
\label{subsec:thermality-Jacobi}

\begin{figure}[htb!]
    \centering
    \includegraphics[width=0.475\linewidth]{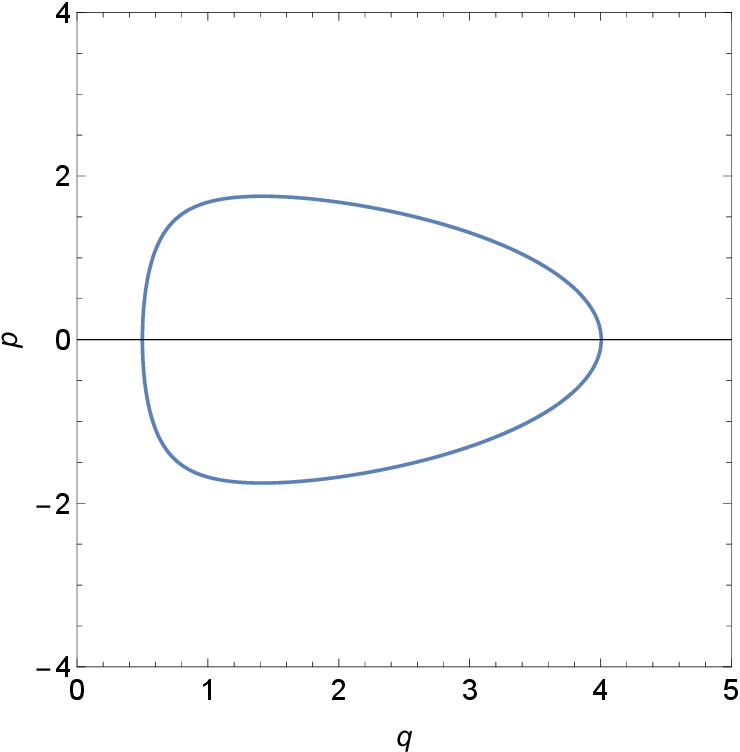}
    \caption{Example of a closed trajectory for $R$ (with $g=1,\alpha=2$, and $E_{\gamma}=2$).}
    \label{fig:R-trajectory}
\end{figure}

We can now reexamine the microcanonical analysis of Sec.~\ref{subsec:thermality-from-duality}
via the explicit evaluation of the Jacobi action,
by considering closed periodic orbits $\gamma$ of $R$ in classical phase space.
As shown in Sec.~\ref{subsec:PI-functionals},
these periodic orbits, despite being classical, are the ones used in
the quantum-mechanical traces~(\ref{eq:trace-K}) and (\ref{eq:trace-G})
 in the path-integral formulation of the theory. As such, they encode the full-fledged quantum and thermal behavior, but the origin of this property can be probed at the semiclassical level.
Moreover, they are the basic building blocks in the study of both classical and quantum chaos.
In dealing with conservative systems, the existence of these orbits is guaranteed, leading to their definition via the Hamiltonian  $R \equiv \tilde{H}_{R}$
(denoted with the symbol of the corresponding generator),
given as in Eq.~(\ref{eq:Hamiltonian-orbits-for-R}).
A typical trajectory of constant energy $E_{\gamma}$ is shown in Fig.~\ref{fig:R-trajectory}.

For the description of the orbits of the operator $R$ defined by Eq.~(\ref{eq:Hamiltonian-orbits-for-R}), a fixed-energy condition is required, and this converts the usual action $S[q;T]$ into the Legendre-transformed Jacobi action~(\ref{eq:Jacobi-action-as-Legendre}).
For the periodic orbits, the Jacobi action integral $W(E_{\gamma})$ 
admits the following geometrical interpretation:
it gives the area enclosed by the closed path in Fig.~\ref{fig:R-trajectory}, which can 
 be expressed in closed form as
\begin{equation}
\begin{aligned}
W(E_{\gamma}) 
= \oint p \, dq  & = 
 \oint \, dq  \, \sqrt{2 E_{\gamma}
 -\frac{g}{q^2} -  \frac{1}{  \alpha_{\!_{R}}^2}q^2
  } 
\\
& =    \pi  \alpha_{\!_{R}}
   E_{\gamma} - \pi \sqrt{g} 
    \; 
\label{eq:action-R}
\end{aligned}
\end{equation}
 (e.g., $2.267.1$ in Ref~\cite{Grads-int}).
The result of Eq.~(\ref{eq:action-R}) shows that the Jacobi action is an increasing
function with respect to $\alpha_{R}$ and decreasing with respect to $g$.

Finally, as Eq.~(\ref{eq:action-R}) has been obtained via a classical argument,
 its correct semiclassical interpretation requires the \textit{Langer correction~\cite{Langer-correction}:
 $\sqrt{g} \rightarrow \sqrt{g+ 1/4} = \mu$\/}, 
when used as part of the phase of a WKB or quasiclassical expansion
in the presence of inverse square potential terms~\cite{nhcamblong,nhcamblong-sc}. 
Thus, the correct semiclassical result (to be used in the next section) is
\begin{equation}
\left.
W(E_{\gamma}) 
\right|_{\rm Langer-corr}
=
  \pi  \alpha_{\!_{R}}
   E_{\gamma} - \pi \mu
    \; .
\label{eq:action-R_Langer-corrected}
\end{equation}

\subsection{Semiclassical analysis: Microcanonical approach and information scrambling}
\label{subsec:microcanonical}

The Langer-corrected Eq.~(\ref{eq:action-R_Langer-corrected})
 allows the computation of the period $T_{\gamma}$
via the general equation for the Legendre-transformed conjugate classical transit 
time~(\ref{eq:time-from-Jacobi-action}).
Thus, the time period $T_{\gamma}$ of these closed orbits is
\begin{equation}
    T_{\gamma} = \frac{\partial W(E_{\gamma})}{\partial E_{\gamma}} = \pi \alpha_{\!_{R}}
    \; ,
    \label{eq:period-from-action_02}
\end{equation}
which agrees with Eq.~(\ref{eq:period-from-action}), 
confirming our conclusions on thermality of Sec.~\ref{sec:semiclassical_GF-thermality}.
In particular, this yields the diamond temperature~(\ref{eq:diamond-temperature}). 

A final point is in order regarding the crucially important relationship of the diamond temperature~(\ref{eq:diamond-temperature})
with the instability analysis of Sec.~\ref{subsec:phase-space_Lyapunov}, along
with the related quantum information scrambling.
The information scrambling rate $\lambda_L$ for causal diamonds, according to 
 Eq.~(\ref{eq:Lyapunov_causal-diamonds}), satisfies the upper-bound condition,
 \begin{equation}
 \lambda_L = \frac{1}{ \alpha}
  < \frac{ 2\pi T_{D}}{ \hbar} = \frac{2}{ \alpha}
\;   ,
\end{equation}
 with a strict inequality.
 This upper bound is determined by the observer's lifetime or diamond size, just as the
upper bound for black holes is determined by the surface gravity.
 However, \textit{unlike the case of black holes, the upper-bound is not saturated.\/}
Most importantly, the main lesson from this analysis is that an instability is instrumental for nontrivial thermal 
and quantum information scrambling effects in spacetime, 
just as it was earlier found in Ref.~\cite{majhi-nh-thermality1} for black holes.

\section{Discussion and further work}
\label{sec:discussion}

We have shown that a finite-lifetime observer in Minkowski spacetime, described within 
a causal diamond, detects a thermal state with temperature $T_D$, Eq.~(\ref{eq:diamond-temperature}).
This prediction is solely inferred 
within a path-integral framework for CQM, from the properties of the time-evolution generator, which is the 
hyperbolic $S$ operator of the dAFF model, confirming properties known from a variety of alternative approaches. Moreover, these path-integral findings further corroborate the role of the effective potentials
including an inverted harmonic oscillator~\cite{IHO-Barton}
 to generate thermal behavior linked to an instability, 
extending them to the remarkable physics of causal diamonds in flat spacetime. 

Even though our findings concerning the relation between the \textit{instability and thermality\/}
have been partially motivated via a semiclassical limit in a manner related to the work of Refs.~\cite{majhi-nh-chaos,majhi-nh-thermality1,majhi-nh-thermality2,majhi-thermality-Kerr,majhi-nh-thermality-ansatz,majhi-nh-thermality-PI},
we have \textit{extended them to a more general, exact quantum-mechanical argument with path integrals,
via analytic continuation with the canonical partition function of the operator $R$.\/}
We are currently investigating methods to establish a more direct connection of quantum mechanical chaos with the $S$ operator using fully quantum mechanical tools such as the finite-temperature out-of-time-order correlation (OTOC) function~\cite{OTOC3,OTOC5}, which can be applied to the single particle operator $S$, 
e.g., following the footsteps of Ref.~\cite{hashimoto}. 

 Moreover, the CQM model is also related to the Sachdev-Ye-Kitaev (SYK) model in the large-$N$ limit~\cite{SYK-model,syk-maldacena}, 
 with connections to black holes in AdS spacetime \cite{chaos-ads2} and applications to a broad range 
 of areas of physics regarding information scrambling~\cite{Qinfo-review_LewisSwan}.
 It seems plausible that further connections of our line of work with the SYK model may provide additional insights into the links between CQM, chaos, and black hole thermodynamics.
  
In closing, the importance of finding experimental realizations of the causal-diamond physics of finite-lifetime observers
should be highlighted. This could potentially lead to critical tests of 
relativistic quantum information. In fact, the possible use of time-dependent Stark or Zeeman effects has been considered~\cite{Su2016SpacetimeDiamonds}, but 
additional phenomenological analysis is needed for its implementation.

\acknowledgments{}
This material is based upon work supported by the Air Force Office of Scientific Research under Grant No. FA9550-21-1-0017 (C.R.O., A.C., and P.L.D.).
C.R.O. was partially supported by the Army Research Office (ARO), grant W911NF-23-1-0202.
H.E.C. acknowledges support by the University of San Francisco Faculty Development Fund.

\begin{appendix}

\section{PATH-INTEGRAL AMPLITUDE FUNCTIONALS: REVIEW OF GENERAL DEFINITIONS}
\label{sec:app_PI-functionals}

In this appendix, we review and adjust without proof the basic definitions and properties
of path integrals. These are given in both their canonical and microcanonical forms, i.e., as the propagators and the energy Green's functions, in a form suitable for the analysis in the main text. 
This summary is formulated with the higher-dimensional expressions that are useful for a proper interpretation
of the CQM functionals with a nontrivial inverse square potential term (as given in the next section).

The definitions, properties, and notation of both appendices are basically modeled after 
Refs.~\cite{PI-CQM,Cam_PI-singular-ISP,Green_operator_approach}
(where the first one deals with all the CQM generators, and the last two deal with CQM without the extra harmonic oscillator potential, but also including the strong coupling regime); in addition, Refs.~\cite{Kleinert-PI,Grosche-PI} provide a more extensive background.
In both appendices, we will keep all the physical dimensions to consider appropriate limits and relationships between different sectors of the theory.

\subsection{Canonical functional integrals}
\label{sec:app_canonical-PIs}

The general expressions below apply to a particle of mass $M$ subject to a Hamiltonian 
$\hat{H}=p^2/2M + V({\bf r})$ in $d$ spatial dimensions, where the potential is
 time-independent and central.
The quantum-mechanical propagator or transition amplitude is given by
\begin{align} 
K_{(d)}^{(H)} ( {\bf r}'', {\bf r}' ; t'',t' ) 
&
=
\left\langle 
{\bf r}''
\left| 
\hat{T}
\exp \left[
 -\frac{i}{\hbar} \int_{t'}^{t''}
\hat{H} dt 
\right]
\right|
{\bf r}'
 \right\rangle   
\label{eq:propagator_QM_from_evolution_operator}
 \\
 &=
\int_{  {\bf r} (t')  = {\bf r}'  }^{  {\bf r} (t'')  = {\bf r}'' }
 \;  
{\cal D} {\bf r} (t) \,
\exp \left\{ 
\frac{i}{\hbar} 
S \left[ {\bf r}(t)  \right]  ({\bf r}'', {\bf r}' ; T)  
\right\}
\; ,
\label{eq:propagator_QM}
\end{align}
where $\hat{T}$ is the time-ordering operator.
As in the main text, $S \left[ {\bf r}(t)  \right]  ({\bf r}'', {\bf r}' ; T)  $  is
the ordinary Lagrangian action functional
for ``paths''
$ {\bf r}(t)  $ connecting the end points
${\bf r} (t')  = {\bf r}',  {\bf r} (t'')  = {\bf r}'' $,
where the action and propagator are
functions of $T=t''-t'$ alone for time-independent potentials.
Equation~(\ref{eq:propagator_QM_from_evolution_operator})
refers to a generic Hamiltonian $\hat{H}$, stating that 
$K_{(d)}^{(H)}  ({\bf r}'', {\bf r}' ; T)  = 
\left\langle 
{\bf r}''
\left| 
U_{H}(T)
\right|
{\bf r}'
 \right\rangle 
 $, with $U_{H}(T)$ being the time-evolution operator displayed above; this is relevant for the discussion
 of CQM generators, as in Appendix~\ref{sec:app_CQM-generators-summary}
 and in the main text, because we are dealing with an infinite set of generators $\hat{G}$
 with associated functionals 
 $K_{(d)}^{(G)}  ({\bf r}'', {\bf r}' ; T)  $.
 
Using a basis of $d$-dimensional hyperspherical harmonics $Y_{l \boldsymbol{m}} ({\bf \Omega})$,  
with hyperspherical polar coordinates ${\bf \Omega}$~\cite{Cam_DT1,Cam_DT2,Erdelyi_HTF2},
 the propagator can be expanded in terms of the building blocks with separated variables  
 ${\bf \Omega}$ and $r$. This reduces the problem~\cite{Kleinert-PI,Grosche-PI,PI-CQM} to 
 a radial propagator $K_{l+\nu}(r'',r';T)$, which is defined via the partial-wave expansion
\begin{equation}
 \begin{aligned}
& K_{(d)}({\bf r}'', {\bf r}' ; T) 
  =
\frac{\Gamma (\nu)}{2 \pi^{d/2} }
\,
\left( r'' r' \right)^{-(d-1)/2}
\,
\sum_{l= 0}^{\infty}
\,
(l+\nu)
\,
C_{l}^{(\nu)} (\cos \psi_{ {\bf \Omega}'', {\bf \Omega}' } ) 
\,
K_{l +\nu}(r'',r';T)
\;  ,
\label{eq:propagator_partial_wave_exp}
\end{aligned}
\end{equation}
where $\nu =d/2 -1$,
 $\cos \psi_{ \, {\bf \Omega}'', {\bf \Omega}' } = {\bf \hat{r}}'' \cdot {\bf \hat{r}}' $
  (with $\hat{\mathbf{r} } = \mathbf{r}/r$), 
  and  $C_{l}^{(\nu)}(x)$ are
Gegenbauer polynomials, to be further expanded into spherical harmonics.
The $d$-dimensional angular momentum is labeled with quantum numbers
$l$ and $\boldsymbol{m}$~\cite{Kleinert-PI,Grosche-PI,Cam_DT2}
(Chap.~XI in Ref.~\cite{Erdelyi_HTF2} and \cite{Kleinert-PI, Grosche-PI}).
For central potentials, the radial propagator $K_{l+\nu}(r'',r';T)$ is independent of the angular coordinates 
and quantum numbers $\boldsymbol{m}$, and it has the form of a standard one-dimensional path-integral
(in the half-line $r(t) \geq 0$), with the same normalization factors, in addition 
to a multiplicative measure factor with Bessel functions of order $l+\nu$~\cite{Kleinert-PI,Grosche-PI,PI-CQM}. 
This factor is used to deal with an inverse square potential as an extra angular momentum term, 
and gives a closed path-integral solution for the CQM generators
in radial form: $K^{(G)}_{l+\nu}(r'',r';T)$, evolving with a corresponding effective time $\tau$---see a partial summary in Appendix~\ref{sec:app_CQM-generators-summary}.

In particular, as the one-dimensional case ($d=1$) is considered in the main text and parts of these appendices, we 
specify the corresponding expansion formulas, in terms of a 1D variable $q$.
Specifically, $\nu = -1/2$, $r=|q|$, and there are two parity angular-momentum channels $l=0,1$ 
(even/odd, $l+\nu = \mp 1/2$ respectively), with polynomials
$C_{l}^{(-1/2)} (\epsilon )  = (-\epsilon)^{l}$, where $\epsilon = {\rm sgn} (q' q'')$.
Then,
 $  K^{}(q'',q';T) =
 [ K^{}_{-1/2}(r'',r';T) + \epsilon K^{}_{1/2}(r'',r';T)]/2 $.
 This setup allows for arbitrary problems where $q$ extends over the entire real line $q \in (-\infty,\infty)$; but, when the problem is restricted to the half-line
 $q \in [0,\infty)$, the partial propagators are the same, $ K^{}_{-1/2} = K^{}_{1/2}$, giving 
 a unique propagator $ K = K^{}_{-1/2} = K^{}_{1/2}$ that vanishes for ${\rm sgn} \, (q'') \neq {\rm  sgn} \, (q')$, which, starting from $q'>0$ restricts the problem to $q''>0$.
   This is what happens in our problem of interest in the presence of a repulsive inverse square potential,
   which drastically cuts off the real line into two disjoint problems.
 
 \vspace{-0.0025in}
 
\subsection{Microcanonical functional integrals}
\label{sec:app_microcanonical-PIs}

For the microcanonical functional integrals, the standard retarded/advanced Green's functions
 are defined from
\begin{align}
G^{(\pm)}({\bf r}'',{\bf r}';E) &]
=  
\pm \frac{1}{i \hbar} \int_{-\infty}^{\infty} d T \, \theta (\pm T)
  \, e^{iET/\hbar}\; K({\bf r}'',{\bf r}';T)
      \label{eq:Green-functions-def}
       \\
   &
    = 
\pm  \frac{1}{i \hbar} \int_{-\infty}^{\infty} d T \, \theta (\pm T)
    \int_{  {q} (t')  = {q}'  }^{  {q} (t'')  = {q}'' }
 \;  
{\cal D} {\bf r} (t) \,
\exp \left\{ 
\frac{i}{\hbar} 
W \left[ {\bf r}(t)  \right]  ({\bf r}'', {\bf r}' ; E)  
\right\}
    \; ,
    \label{eq:Green-functions-microcanonical}
        \end{align}
where $\theta $ stands for the Heaviside function, the replacement 
 $E \rightarrow E \pm i 0^{+} $ guarantees convergence, and
 $W \left[ {\bf r}(t)  \right]  ({\bf r}'', {q}' ; E)  $ in the exponent is the Jacobi action
 defined as in Eq.~(\ref{eq:Jacobi-action-as-Legendre}).
 The definitions~(\ref{eq:Green-functions-def}) involve the Fourier transforms of the retarded/advanced 
 time Green's functions, which correspond to  the (retarded/advanced) Green's operators
$
\hat{G}^{(\pm)} (T) =  \theta (\pm T) \, e^{-i \hat{H} T/\hbar}
$,
leading to the energy Green's operators
\begin{equation}
G^{(\pm)} (E)= \left( E - \hat{H} \pm i 0^{+} \right)^{-1}
\; 
\label{eq:Green-operators-energy}
\end{equation}
[with extra factors $\pm (1/i\hbar)$], 
where the $i 0^{+}$ prescription provides convergence for each case. 
 These equations for the Green's functions can be applied with the same structural form
 either to the full-fledged multidimensional quantities $G^{(\pm)}({\bf r}'',{\bf r}';E) $, as in Eq.~(\ref{eq:Green-functions-def}),
 or to their reduced radial counterparts
 $G_{l+\nu} (r'',r';E) $,
 related to $G^{(\pm)}({\bf r}'',{\bf r}';E) $
 via the same hyperspherical expansion~(\ref{eq:propagator_partial_wave_exp}) that defines the radial propagators.

As a consequence of the definitions of the Green's functions, 
via the identity 
$\left( A \pm i 0^{+} \right) ^{-1} = \mathcal{P}(A^{-1}) \mp i \pi \delta (A)$ (with 
$\mathcal{P}$ being the Cauchy principal value),
the density of states $\rho (E)$ can be computed.
The practical formulas for $\rho (E)$ take the form
\begin{equation}
  \begin{aligned}
      \rho(E) &
      = -\frac{1}{2 \pi i} {\rm Tr} \left[  \hat{G}^{(+)}(E)  -  \hat{G}^{(-)}(E)  \right]
      = -\frac{1}{\pi} {\rm Im} {\rm Tr} \left[  \hat{G}^{(+)}(E)   \right]
    \\
    & =
    -\frac{1}{\pi} {\rm Im}\left[\int d^{d} r \; G^{(+)}(q,q;E) \right]
    \end{aligned}  
    \; ,
    \label{eq:dos-expression}
\end{equation}
where in the last line the trace is computed in the configuration-space representation. 
(For the radial Green's functions, this is just an integral
 $\displaystyle \int_{0}^{\infty} dr$.)

In the following discussion, beyond this introductory review, we will simply
use the retarded Green's function, with the simplified notation $G \equiv G^{(+)}$.

\section{PROPERTIES OF FUNCTIONAL INTEGRALS OF CQM GENERATORS: SUMMARY AND ANALYTIC CONTINUATIONS}
\label{sec:app_CQM-generators-summary}

In this appendix, we reexamine the path integrals of the CQM generators,
including propagators and Green's functions; and we formulate
the analytic relations between the elliptic and hyperbolic operators, i.e., between $R$ and $S$.
 The basic expressions were fully derived in Ref.~\cite{PI-CQM}
 for a generalized dAFF model in any number of dimensions.
 For the discussion below,
 it will also be useful to keep all the physical parameters,
 including mass $M$ and $\hbar$ as in Ref.~\cite{PI-CQM}.
 
 \subsection{Basic elliptic and hyperbolic path integrals of CQM}
 \label{sec:app_basic-PI-CQM}
 
 First, there are only
 three basic path integrals, corresponding to each of the three classes of generators defined from the 
 discriminant~(\ref{eq:gen-generator_discriminant}):
 elliptic ($\Delta <0$), parabolic ($\Delta =0$), and hyperbolic ($\Delta >0$).
 We can refer to these as just $R$, $H$, and $S$, respectively, but the path integrals have an identical 
 form for all generators of the same class.
 These can also be interpreted with
the analog quantum-mechanical problem as in Eq.~(\ref{eq:potential_HO+ISP}), with a Hamiltonian
\begin{equation}
 \tilde{H} (r) = \frac{p^2}{2M}
 + \frac{1}{2} M \omega^2 r^2 + \frac{\hbar^2}{2M} \, \frac{g}{r^{2}}
\label{eq:Hamiltonian_HO+ISP}
 \; .
\end{equation}
  Moreover, $S$ and $R$ are related by a simple analytic continuation, while $H$ can be regarded as a limiting case; thus, one can start with $R$, which corresponds to an ordinary harmonic oscillator combined with an inverse square potential, and get the others by appropriate extensions.
 For our purposes, we will only focus on $R$ and $S$ below.

 Second, the primary results for the propagators 
will be stated in their radial forms for CQM 
in $d$ conformal field variables
[field $Q$ with $d$ components in Eq.~(\ref{eq:CQM_action})
interpreted as a $d$-dimensional set].
Then, the relevant radial dependence from the conformal inverse square potential
is parametrized with the conformal index 
\begin{equation}
\mu = \sqrt{ (l + \nu)^2 + g}
\; ,
\end{equation}
 where $g$ is the conformal coupling, and
$\nu = d/2-1$ in $d$ spatial dimensions. 
(The label of the functionals refers to the interdimensional angular momentum variable $l + \nu$.)

Third, the primary expressions are as follows.
For the elliptic operator, the general propagator is 
\begin{equation} 
    K_{l+\nu}^{(R)} (r'',r';T) 
=
\frac{M \omega }{i \hbar \sin \omega T} 
\,
\sqrt{ r' r''} 
\,
\exp \left[ \frac{i M \omega }{2 \hbar } \left(r'^{2} + r''^{2} \right) \cot \omega T \right]
I_{\mu} \left( \frac{M \omega r'r''}{i \hbar \sin \omega T} \right)
\;   ,
\label{eq:propagator_R}
\end{equation}
where $I_{\mu}(z)$ is the modified Bessel function of the first kind and order $\mu$.
This expression is completely general for complex values of the parameter $\omega$; thus, it can be used to relate 
the elliptic and hyperbolic sectors of the theory, using
  \begin{equation}
    \left\{
    \begin{array}{c}
    R \longrightarrow S
        \\
    S  \longrightarrow R
        \end{array}
    \right\}
     =
    \begin{pmatrix}
   \omega \longrightarrow \mp i \omega
         \end{pmatrix}
         \label{eq:S-to-R_omega} 
         \; ;
   \end{equation}
this is the same as the duality of Eq.~(\ref{eq:S-to-R_alpha}). 
Thus, 
 with the replacement $\omega\rightarrow - i\omega$, Eq.~(\ref{eq:propagator_R})
 gives the propagator for hyperbolic operators,
\begin{equation}
    K_{l+\nu}^{(S)} (r'',r';T) 
    =
     \frac{M \omega}{i  \hbar \sinh \omega T } \sqrt{r'r''}
\,  \exp \left[  \frac{ i M \omega}{2 \hbar}  \left( {r'}^{2}+{r''}^{2} \right)  \coth \omega T \right]
\,
I_{\mu}  \left(\frac{ M \omega r'r''}{i \hbar \sinh \omega T} \right)  
\; .
\label{eq:propagator_S}
  \end{equation}
Moreover, as mentioned in the main text,
 the choice $\omega\rightarrow- i\omega$ (instead of the opposite sign)
has the operational advantage that it preserves the boundary condition at infinity
corresponding to retarded Green's functions~\cite{PI-CQM}, and it is also
consistent with the standard extension to Euclidean time (see below).
In addition, Eq.~(\ref{eq:S-to-R_omega}) can also be applied to the Green's functions, e.g., 
$ \left. G_{l+\nu}^{(R)} (r'',r';E) 
\right|_{\omega\rightarrow- i\omega} 
=
G_{l+\nu}^{(S)} (r'',r';E) $.
Thus, it suffices to specify, as shown in Ref.~\cite{PI-CQM},
  \begin{equation}
  G^{(\pm)(R)}_{l+\nu} (r'',r';E) =
  \mp     ( \hbar \omega )^{-1}
 \, 
\frac{
 \Gamma \left( (1+\mu)/2 \mp  \kappa \right)
}{\Gamma (1+\mu)}\,   \frac{1}{ \sqrt{ r' r''  } }
 W_{\pm \kappa,\mu/2} ( \pm  \check{r}_{>}^{2} )   
   \,  {M}_{ \pm \kappa,\mu/2}   ( \pm  \check{r}_{<}^{2} )
\; ,
\label{eq:energy-Green-functions_R}
  \end{equation}
  where
  $ {M}_{\lambda,\mu/2}  (z)$
and
$W_{\lambda,\mu/2} (z)  $ are Whittaker functions~\cite{Grads-int},
$\kappa  = \tilde{E}/2 \hbar \omega$ [for the eigenvalues of the operator~(\ref{eq:Hamiltonian_HO+ISP})],
  $\displaystyle \check{r}^{2} =   \frac{M  \omega}{\hbar} \,   {r}^{2}$,
 and $r_{>}$  and $r_{<}$
are the greater and lesser of the set $\{r', r''\}$.

Fourth, the simplest realization of this $d$-dimensional theory is the 
one-dimensional case ($d=1$, i.e., $\nu = - 1/2$), as introduced in the original dAFF model and used in our current paper. 
Due to the presence of a nonzero repulsive inverse square potential, with $g >0$,
the one-dimensional problem is restricted to a half-line axis ($r \geq 0$, with $r$ labeled as $q$ in the main body of this work.) 
This choice corresponds to the following set of radial propagators and indices:
\begin{equation}
 K^{(R)}_{l+\nu}(r'',r';T) =K^{(R)}_{\mp 1/2}(r'',r';T)
 \; , \; \; \text{ with } \;  l+\nu = \mp 1/2  \; \; \; \text{ and } \;   \mu = \sqrt{g + 1/4}
 \; ,
 \label{eq:1D-R-propagator}
 \end{equation}
which includes the two angular momentum channels $l=0,1$ with equal contributions.
This procedure is enforced via the Bessel functions with the common positive square root for the conformal index 
 $ \mu = \sqrt{g + 1/4}>0$, as opposed to the distinct indices $\mu = \pm 1/2$ that would be used for a pure oscillator with $g=0$.
As discussed in Appendix~\ref{sec:app_PI-functionals}, Sec.~\ref{sec:app_canonical-PIs},
with the equality of the propagators in Eq.~(\ref{eq:1D-R-propagator}),
the partial-wave expansion gives the same function 
 $ K = K^{}_{-1/2} = K^{}_{1/2}$
 for all the propagators, and similarly, for all the Green's functions,
 $ G= G^{}_{-1/2} = G^{}_{1/2}$.
 In hindsight, the restriction $q>0$ could be handled
 bypassing some of these details, by choosing $K^{(R)}_{1/2}(r'',r';T)$ 
 (odd-channel,
 $l=1$, solutions)
 with $\mu = \sqrt{g + 1/4}$
 as the main propagator from scratch,
 according to Eq.~(\ref{eq:1D-R-propagator}).
 
Finally, another analytic continuation, central to our derivations, is the extension to imaginary time
via the Wick rotation~\cite{Wick-rotation}
\begin{equation}
\begin{pmatrix}
\text{Real-time functional}
\\
\text{with} \; T \longrightarrow -iT
\end{pmatrix}
=
\begin{pmatrix}
\text{Euclidean}
\\
\text{functional}
\end{pmatrix}
\; .
\label{eq:Euclideanization} 
\end{equation}
Thus, as derived in Ref.~\cite{PI-CQM}, from Eq.~(\ref{eq:propagator_R}), the corresponding Euclidean-time path integral is
 \begin{equation}
 \begin{aligned}
\!  \!  \!  \!  
 K^{(R_{\rm Eucl})}_{l+\nu} (r'',r';T) 
 &  \equiv K^{(R)}_{l+\nu} (r'',r';-i T) 
\\
& =
\frac{M \omega }{ \hbar \sinh \omega T} 
\,
\sqrt{ r' r''} 
\,
\exp \left[ - \frac{M \omega }{2 \hbar } \left(r'^{2} + r''^{2} \right) \coth \omega T \right]
I_{\mu} \left( \frac{M \omega r'r''}{ \hbar \sinh \omega T} \right)
\,   .
\label{eq:propagator_R_Euclidean}
\end{aligned}
\end{equation}

\subsection{Relationships between elliptic and hyperbolic path integrals of CQM}
\label{sec:app_elliptic-hyperbolic_PI-relations}

The propagators~(\ref{eq:propagator_R})--(\ref{eq:propagator_R_Euclidean}),
their corresponding traces, and the associated Green's functions satisfy simple equivalence relations that can be derived from first principles and/or the explicit form of these functional integrals.
These relations provide an important generic tool in our analysis of CQM:
the ability to relate the two nontrivial sectors of the theory, elliptic and hyperbolic
[beyond the original Hamiltonian $H$ corresponding to Eq.~(\ref{eq:CQM_action})].
This is based on the reasonable ansatz that the Euclidean-time version of a theory with a potential generates a
problem in imaginary time with the inverted potential. In the case of CQM, when the inverse square potential is
treated separately as part of the functional measure (or absorbed in a nontrivial angular momentum), the inversion is only enforced for the harmonic oscillator, which amounts to the analytic continuation 
$\omega\rightarrow - i\omega$,
i.e., the conversion of $R$ into $S$.
Thus, barring minor adjustments, there should exist relations of the form
\begin{equation}
\begin{pmatrix}
\text{Functional}
\\
\text{for} \; R_{\rm Eucl}
\end{pmatrix}
\sim
\begin{pmatrix}
\text{Functional}
\\
\text{for} \; S
\end{pmatrix}
\; .
\label{eq:relations_R-to-S}
\end{equation}
In short, the {\em network of analytic-continuation relations among the functionals\/} consists of
Eqs.~(\ref{eq:S-to-R_omega}),
(\ref{eq:Euclideanization}),
and
(\ref{eq:relations_R-to-S}).
However, we have to specify what is meant by the subset of relations~(\ref{eq:relations_R-to-S}).
These include the following statements, which we prove below.
\begin{itemize}
\item Propagators:
\begin{equation}
   \left.
   K^{(R_{\rm Eucl})}_{l+\nu} (r'',r';T) 
    \right|_{M \rightarrow - i M}
   =
      K^{(S)}_{l+\nu} (r'',r';T) 
\; .
\label{eq:propagator-relation_R-to-S}
  \end{equation}

\item Energy Green's functions:
\begin{equation}
  \left.
   G^{(R_{\rm Eucl})}_{l+\nu} (r'',r';E) 
     \right|_{\omega \rightarrow - i \omega}
   =
   G^{(S)}_{l+\nu} (r'',r';E) 
\; .
\label{eq:GF-relation_R-to-S}
  \end{equation}

\item Propagator traces:
\begin{align}
  \tilde{Z}_{l+\nu} ^{(R_{\rm Eucl})}(T) 
& =  \tilde{Z}_{l+\nu} ^{(S)}(T)
\label{eq:propagator-trace-relation_R-to-S}
\\
& =
\frac{e^{-\mu \omega T}}{2\sinh \omega T}
\; .
\label{eq:propagator-trace-relation_explicit}
  \end{align}

\item Energy Green's function traces:
\begin{equation}
  \left.
  {\rm Tr} \left[ \hat{G}^{(R_{\rm Eucl})} (E)  \right]
     \right|_{\omega \rightarrow - i \omega}
 =
   {\rm Tr} \left[ \hat{G}^{(S)}(E)  \right]
  \; .
    \label{eq:GF-trace-relation_R-to-S}
  \end{equation}

\end{itemize}

The first basic relation~(\ref{eq:propagator-relation_R-to-S})
 is the direct formal connection between the propagators.
 Its validity is due to that the full-fledged real-time evolution of $S$ can be restored with an imaginary mass,
which multiplies all the terms of the Hamiltonian
in Eq.~(\ref{eq:Hamiltonian_HO+ISP}). This is also 
explicitly verified from a direct comparison of the resulting propagators~(\ref{eq:propagator_S}) and 
(\ref{eq:propagator_R_Euclidean}).

The second relation~(\ref{eq:GF-relation_R-to-S})
is a similar formal connection between the energy Green's functions.
This is due to the replacement of the time variable by the energy $E$; and by definition,
the same energy Green's function is introduced in either framework
in such a way that the Green's operators take the form~(\ref{eq:Green-operators-energy}).
Thus, the analytic continuation simply reduces back to Eq.~(\ref{eq:S-to-R_omega}).
This is also explicitly verified with the form of the energy Green's
functions~(\ref{eq:energy-Green-functions_R})---with additional details in Ref.~\cite{PI-CQM}.

The fourth relation~(\ref{eq:GF-trace-relation_R-to-S})
is a consequence of the identity of energy Green's functions, Eq.~(\ref{eq:GF-relation_R-to-S}).
This directly leads to the formal identity of the corresponding density of states,
but their values need to be adjusted as follows.
The formal connection between both relations for the energy Green's functions: 
Eqs.~(\ref{eq:GF-relation_R-to-S}) and (\ref{eq:GF-trace-relation_R-to-S}),
involve the subtle replacement of the frequency parameter required by the analytic continuation;
in particular, this rotation in the frequency complex plane changes the nature of the functions, making the density
of states~(\ref{eq:GF-trace-relation_R-to-S}) divergent, in a way that requires regularization 
(see Appendix~\ref{sec:app_DOS_operator-S}).

Finally, the third relation, 
Eqs.~(\ref{eq:propagator-trace-relation_R-to-S})--(\ref{eq:propagator-trace-relation_explicit})
requires a more detailed analysis.
This central identity is the key to the proof of the characterization of thermality,
Eq.~(\ref{eq:partition-function-S_01}).
This is the statement of identity of the traces, when defined via Eq.~(\ref{eq:trace-K}), 
and can be proved as follows.
First, without finding its specific value,
the equality
$\tilde{Z}_{l+\nu} ^{(R_{\rm Eucl})}(T) = \tilde{Z}_{l+\nu} ^{(S)}(T)$ is based on the scaling 
of the original Hamiltonian with the variable $M r^2$, as can be seen from 
Eq.~(\ref{eq:Hamiltonian_HO+ISP}). This scaling corresponds to the analytic extension $M \longrightarrow -iM$, which
relates the two propagators; then, when computing the corresponding traces, which are radial integrals, the final outcome only depends on the combination $Mr^2/\hbar$, such that the extension $ M r^2 \longrightarrow -iMr^2$
converts $\tilde{Z}_{l+\nu} ^{(R_{\rm Eucl})}(T)$ exactly into $\tilde{Z}_{l+\nu} ^{(S)}(T)$.
This common value, computed as $\tilde{Z}_{l+\nu} (T)$ with the operator $S$, can be rewritten with 
the substitution $
x ={M\omega r^2}/{\hbar}$, yielding
\begin{equation}
\tilde{Z}^{(S)}(T)
=
{\rm Tr }
\left[ K_{l+\nu}^{(S)} \right]
=
\int_0^\infty 
dr\; K_{l + \nu}^{(S)}(r,r;T)
= \frac{1}{2i \sinh{\omega T}}
 \int_0^\infty dx \;e^{ix\coth \omega T} I_\mu\left(\frac{x}{i\sinh{\omega T}}
 \right) 
 \; .
\label{eq:Tr-K-S-int}
\end{equation}
Then, in Eq.~(\ref{eq:Tr-K-S-int}), with $z=\omega T$,
the integral 
\begin{equation}
  \mathcal{I} (z,\mu) =  
   \int_0^{\infty} e^{ix\coth z} \; I_{\mu}\left(\frac{x}{i\sinh z}\right) \; dx 
  = i
  \,  \int_0^{\infty} e^{-y\coth z} \; I_{\mu}\left(\frac{y}{\sinh z}\right) \; dy
   = i \;e^{-\mu z} 
   \label{eq:partitionZ_aux-integral}
   \end{equation}
is evaluated using either one of the equations~$6.611.1$ 
or $6.622.3$ in Ref~\cite{Grads-int};
for the latter, with the replacements $\zeta =y/\sinh z$ and $a=z$,
the more general integral in Ref~\cite{Grads-int},
\begin{equation*} 
    \int_0^{\infty} 
    e^{-\zeta \cosh a }I_{\mu}(\zeta) \, \zeta^{\nu-1} \, d \zeta
    =\sqrt{\frac{2}{\pi}} \, e^{-(\nu-1/2)\pi i} 
    \; \frac{ Q_{\mu-\frac{1}{2}}^{\nu-\frac{1}{2}} \bigl( \cosh a \bigr) }{ \sinh^{\nu-\frac{1}{2}} a }
     \;,
\end{equation*}
involves  the associated Legendre function of the second kind $Q_{\mu}^{\nu}$,
with $Q^{1/2}_{\mu - 1/2}(\cosh a )= i \, e^{-\mu a} \, \sqrt{\pi/2 \sinh a} $
(Ref~\cite{Grads-int}, Secs.~8.7--8.8).
[The integrals are quoted in terms of $y$ with the real exponential in standard references, 
as in Ref.~\cite{Grads-int}; the substitution $x=iy$ for the integral with respect to $x$ 
in Eq.~(\ref{eq:partitionZ_aux-integral}) is valid, as it amounts to a contour integral in the lower-right quadrant 
of the complex plane $y$, where the integrand is exponentially small at infinity.]
Thus, with the auxiliary equation~(\ref{eq:partitionZ_aux-integral}),
the integral in Eq.~(\ref{eq:Tr-K-S-int}) gives the final form of the trace,
\begin{equation}
  \tilde{Z}_{l+\nu} ^{(S)}(T)
=
\frac{e^{-\mu \omega T}}{2\sinh \omega T}
\; ,
\label{eq:propagator-trace_S}
\end{equation}
 in agreement with Eq.~(\ref{eq:propagator-trace-relation_explicit}).
 Similarly, for the Euclideanized $R$ operator,
 setting $x=M\omega r^2/\hbar$, 
\begin{equation}
 \tilde{Z}_{l+\nu} ^{(R_{\rm Eucl})}(T) 
=  
    \frac{1}{2\sinh{\omega T}}\;
\int_0^{\infty} dx
 \; \exp\left(
 {-x\coth{\omega T}}
 \right)
 \; 
 I_{\mu}\left(\frac{x}{\sinh{\omega T}}\right) 
 \; , 
\end{equation}
 where, using again the auxiliary equation~(\ref{eq:partitionZ_aux-integral}),
 the partition function becomes
 \begin{equation}
  \tilde{Z}_{l+\nu} ^{(R_{\rm Eucl})}(T) 
      =
            \frac{e^{-\mu \omega {T}}}{2\sinh{\omega {T}}} 
    \; ,
    \label{eq:propagator-trace_R-Eucl}
\end{equation}
which gives the same value as
$  \tilde{Z}_{l+\nu} ^{(S)}(T) $.
This completes the proof of 
Eqs.~(\ref{eq:propagator-trace-relation_R-to-S})--(\ref{eq:propagator-trace-relation_explicit}).
 
Additional path integral results regarding the density of states from the energy Green's functions are discussed in 
the next section.

\section{\mbox{DENSITY OF STATES ASSOCIATED WITH THE OPERATOR \boldmath$S$}}
\label{sec:app_DOS_operator-S}

This appendix develops all the basic results on the density of states $\rho^{(S)}(E) $
corresponding to the operator $S$.
In addition to deriving exact, fully quantum-mechanical expressions for $\rho^{(S)}(E) $,
we highlight various results involving regularization of the resulting 
divergent equations as well as the semiclassical limit. 

These miscellaneous results provide additional insight into the main properties
of thermality and instability of causal diamonds derived in the main text, and can be used to reexamine other
treatments of the problem.

 \subsection{Density of states from the exact energy Green's functions}
 \label{sec:app_DOS_exact-GFs}

For the energy Green's functions, defined to satisfy the resolvent operator 
equation~(\ref{eq:Green-operators-energy}), their trace is
\begin{align}
 {\rm Tr} \left[ \hat{G}_{l+\nu}^{(S)}(E)  \right]
&
=
\frac{1}{i \hbar} \int_{0}^{\infty} d T 
  \, e^{iET/\hbar}\; \tilde{Z}_{l+\nu}^{(S)} (T)  
  =
    \frac{1}{i \hbar}
    \int_0^{\infty} \, 
    dT \; e^{iET/\hbar} \; 
    \frac{e^{-\mu \omega T}}{2\sinh \omega T}
  \\
  &
     =
     \frac{1}{i \hbar \omega}
\int_0^{\infty} \, 
    d { z } \; 
    e^{iE { z }/\hbar\omega} \; 
    \frac{e^{-\mu { z }}}{2\sinh { z }}
    \label{eq:Tr-G-S-int}
    \; 
\end{align}
(assuming the retarded operators/functions with the plus sign),
where ${ z }=\omega T$ and 
Eqs.~(\ref{eq:trace-K})--(\ref{eq:trace-G}) and (\ref{eq:propagator-trace-relation_explicit}) were used.

In what follows, we will consider only the density of states expressions for the original dAFF model, which corresponds to an effective one-dimensional problem ($d=1$).
In that case, from Eq.~(\ref{eq:1D-R-propagator}), 
the conformal index is $ \mu = \sqrt{  g+ 1/4}$.
Also, as discussed in Appendix~\ref{sec:app_PI-functionals}, Sec.~\ref{sec:app_canonical-PIs},
the net path-integral functionals for an effective one-dimensional 
problem restricted to the positive half-line by an inverse square potential are the same as their individual angular momentum
components. In particular,
 $  \hat{G}^{}(E) =     \hat{G}_{l+\nu}^{}(E) $, so that
the density of states is obtained from the retarded Green's functions 
via Eq.~(\ref{eq:dos-expression}).
Then, with Eq.~(\ref{eq:Tr-G-S-int}), the density of states for the operator $S$ has the form
\begin{equation}
  \! \! 
      \rho^{(S)}(E) 
     \!   = 
    -
    \frac{1}{2 \pi \hbar } 
    {\rm Im} \!
    \left[
    \frac{1}{i  \omega} 
    \int_0^{\infty} d{ z }
    \; e^{iE{ z }/\hbar\omega}
     \;\frac{e^{-\mu { z }}}{\sinh{{ z }}}
      \right]
    \!   =  \!
          \frac{1}{2 \pi \hbar} 
{\rm Re} \!
   \left(
   \frac{1}{\omega} 
    \int_0^{\infty} d{ z }
     \;  
     \frac{
     \exp \left[ i (\eta + i \mu )
      { z }  \right]
      }{\sinh{{ z }}}
   \right)
  \, ,       \label{eq:dos_S-integral_02} 
\end{equation}
where $\eta=E/(\hbar \omega)$.
In addition, according to the relation~(\ref{eq:GF-trace-relation_R-to-S}),
this density function 
can also be computed with the Euclideanized version of the $R$ operator,
giving exactly the same result, but it requires \textit{taking the real (or imaginary) part 
 in Eq.~(\ref{eq:dos_S-integral_02}) after the analytic continuation is performed\/}. 

Several useful properties of Eq.~(\ref{eq:dos_S-integral_02}) can be highlighted with the following procedure.
First, the integral
$I= \int dz F(z)$, where $F(z)= e^{i\zeta z} \, (\sinh z)^{-1}$, with $\zeta = \eta + i \mu$, can be evaluated as follows.
Expanding the denominator via a geometric series of exponentials, 
 $F(z)= 2 \sum_{n=0}^{\infty} e^{-z \left[ (1+2n) - i \zeta \right]}$, and integrating,
 Eq.~(\ref{eq:dos_S-integral_02}) yields
\begin{equation}
    \rho^{(S)}(E) 
    = 
           \frac{1}{\pi \hbar} 
           \,
{\rm Re}
   \left(
  \frac{1}{\omega} 
  \sum_{n=0}^{\infty}
  \frac{1}{ \left[
   \left( 2 n + \mu + 1 \right) - i \eta   \right]}
   \right)  
  \; . 
        \label{eq:dos_S-integral_03} 
\end{equation}

The series~(\ref{eq:dos_S-integral_03}) explicitly displays the poles in the denominator, 
with respect to the energy parameter $\eta$, which occur at 
$\eta= - i(2n+\mu +1)$; with a rotation in the omega plane, 
$\omega \longrightarrow i \omega $, these become the discrete eigenvalues of the $R$ 
operator, Eq.~(\ref{eq:partition-function_eigenvalues}).
These poles, which arise from the Fourier transform of the partition 
function, Eqs.~(\ref{eq:partition-function-S_01-b}) and (\ref{eq:propagator-trace-relation_explicit}),
effectively generate the density of states.
Equation~(\ref{eq:dos_S-integral_03})
is a restricted form (via the real part) of the analytic continuation of the
microcanonical counterpart of the familiar canonical energy
expansion of Eq.~(\ref{eq:partition-function-S_02}).

 \subsection{Regularization and semiclassical limit of the density of states}
\label{sec:app_regularization-SC-DOS}
 
Another important property of Eq.~(\ref{eq:dos_S-integral_03}) for our discussion below 
is that it is divergent, thus requiring regularization.
This is a serious technical problem of the density of states of the operator $S$, unlike the original amplitude functions 
(propagator and Green's functions), which results from the trace operation applied to the energy Green's function.
In effect, in Eq.~(\ref{eq:dos_S-integral_03}), taking $\omega$ to be real in this parametrization, the series 
is divergent, in a manner similar to a harmonic series. 
 This should not be surprising because the operator $S$ is not of trace class~\cite{operators_Conway}.
 Thus, all the expressions at the level of $ \rho^{(S)}(E) $ are purely formal, and they require 
 some form of regularization via a careful redefinition of the model. 
 
Some hints of how regularization works for this problem can be revealed from the semiclassical approximation, 
related to various aspects of the analysis of Sec.~\ref{subsec:thermality-from-duality}.
As a first step, by using the corresponding expression for the operator $R$, i.e., via $ \hat{G}^{(R_{\rm Eucl})} (E) $, 
the density of states is well-defined (as for a radial harmonic oscillator in the quantum-mechanical interpretation); this is easily verified by replacing $\omega \longrightarrow i \omega $
in Eq.~(\ref{eq:dos_S-integral_03}) and then taking the real part.
Now, the known classical limit of the density of states can be derived from integration in classical
phase space, and is 
\begin{equation}
\displaystyle  \overline{\rho}(E) 
= 
\frac{1}{  \pi \hbar }
\,
 \int_{q_{-}}^{q_{+}} \,  
 \frac{dq}{ \displaystyle \bigl| p(q;E) \! /M \bigr|   }
 \; ,
 \label{eq:1D-Thomas-Fermi}
 \end{equation}
 where $p(q;E)$ is defined via Eq.~(\ref{eq:Hamiltonian-orbits-for-R}), 
and the integral is between the turning points $q_{\mp}$.
This is known as the Thomas-Fermi approximation, average, or Weyl term 
in the density of states~\cite{Kleinert-PI,Grosche-PI};
and it is completely general within the semiclassical approximation.

 In addition, for an operator
with classical periodic orbits of period $T(E)$, one can write 
  \begin{equation}
  T(E) = 2 
  \int_{q_{-}}^{q_{+}} \,  
 \frac{dq}{ \displaystyle \bigl| p(q;E) \! /M \bigr|   }
 \; , \; \; \text{so that} \;   \;   \; 
  \overline{\rho}(E) 
=
\frac{1}{2 \pi \hbar }
\,
T(E)
\; .
 \label{eq:1D-period-DOS}
 \end{equation}
The final results for the operator $R$ are thus obtained by straightforward integration, which gives  
  $\displaystyle  \overline{\rho}(E) = 1/(2 \hbar \omega)$ 
and $T(E) = \pi/\omega$.
It is noteworthy that these expressions are identical to Eqs.~(\ref{eq:1D-period})--(\ref{eq:period-from-action}),
which are central to our characterization of the thermal nature of causal diamonds.

On the other hand, for the operator $S$, which has a potential with an inverted harmonic oscillator term, 
there is only one finite turning point (and the other limit is at infinity), and this causes a divergence 
in the density of states and transit time.
Using these classical integral expressions, 
an obvious regularization procedure is {\em real-space cutoff regularization\/}, i.e., for a particle confined in a box of length $L$.
Then, by direct integration in the classical limit of the operator $S$,
one gets a logarithmic leading order 
\begin{equation}
\displaystyle  \overline{\rho}^{(S)}(E)  \approx
 \frac{1}{\pi \hbar \omega}
\ln \left( 
\sqrt{\frac{2M}{|E|}} \, \omega L
 \right)
\; ,
\label{eq:Thomas-Fermi-DOS_S-op}
\end{equation}
which is the leading high-energy approximation, when $E \gg \hbar \omega$.
This regularization technique has been used for the pure inverted harmonic oscillator~\cite{IHO-Barton}.

Another powerful approach available for this problem, given the form of the series~(\ref{eq:dos_S-integral_03}),
is a {\em digamma-function regularization\/} technique, which is usually stated as 
the replacement via the formal identity $\sigma (z) \equiv \sum_{n=0}^{\infty} 1/(n+z) = - \psi (z)$, 
where $\psi(z) \equiv \psi^{(0)}(z)$ is the 
digamma (psi) function~\cite{Grads-int,NIST:2010}. This procedure is based on the 
more basic formal identity (Eq.~$5.7.6$ in Ref~\cite{NIST:2010}): $\sigma (z) \equiv \sum_{n=0}^{\infty} 1/(n+z) = - \psi (z) + C$,
where $C= -\gamma + \sum_{n=0}^{\infty} 1/(n+1)$ is an infinite constant 
(including the finite Euler-Mascheroni constant $\gamma$), which can either be assumed to be removed by 
regularization subtraction or interpreted via a physical regularization technique such as the cutoff procedure mentioned above.
Thus, Eq.~(\ref{eq:dos_S-integral_03}) becomes
\begin{equation}
\displaystyle {\rho}^{(S)}(E) 
 =
- \frac{1}{2\pi \hbar }
           \,
{\rm Re}
\biggl(
  \frac{1}{\omega} 
  \,
   \bigl[
\psi \bigl( (\mu+1-i \eta)/2 \bigr) 
- C
\bigr]
\biggr)
\; ,
\label{eq:digamma-DOS_S-op}
\end{equation}
defined up to an energy-independent constant $C$.
While the details are beyond the scope of this work, these considerations give an idea of how to deal with the divergence 
problem, and Eq.~(\ref{eq:digamma-DOS_S-op}), properly interpreted, describes the physical density of states. 
 In a more detailed treatment, one can use asymptotic methods, 
 with a Dirichlet boundary condition at $q=L$, to further interpret and determine the leading behavior 
as $L \rightarrow \infty$, for all energies $E$, and correspondingly adjust the path-integral propagators and Green's functions
derived in Ref.~\cite{PI-CQM}.
One simple check of the correctness of this procedure is to extract the asymptotic limit 
$\eta \gg 1$ (i.e. $E \gg \hbar \omega$) using Stirling's series for the digamma function
(Eq.~$5.11.2$ in Ref~\cite{NIST:2010}): $\psi(z) \sim \ln z -1/(2z) $, 
with $z= (1-i \zeta )/2 \sim -i \eta/2$, whose real part in
Eq.~(\ref{eq:digamma-DOS_S-op})
yields
$\displaystyle {\rho}^{(S)}(E)  \sim
\left[ 1/(2 \pi \hbar \omega) \right]
 \ln (2/|\eta|)$,
 which agrees with Eq.~(\ref{eq:Thomas-Fermi-DOS_S-op}) and 
 fixes the ad-hoc constant via the cutoff-length $L$, 
 with $C= - \ln [\hbar/(M\omega L^2)]$.

 \subsection{Density of states: Contribution from poles and semiclassical regime}
\label{sec:app_DOS-poles-SC}

Even though Eq.~(\ref{eq:dos_S-integral_02}) is a formal expression that requires regularization,
additional information can also be gathered from examination of the integral 
in the complex time $T$ plane.
In effect, in terms of $z=\omega T$,
 the integrand $F({ z })$ has poles on the imaginary axis,
 located at ${ z }_{k} = i \pi k$, where $k$ is an integer,
  with residues 
$R_{k} = (-1)^k e^{i  \zeta  { z }_{k}}
 =
 (-  e^{- \pi \zeta}  )^k 
 = (-  e^{-i \pi \mu} e^{- \pi \eta}  )^k 
$.
These poles give a sort of resonant contribution to the density in the form of a series of terms,
with each one due to a conjugate point of the action in the path integral (in the stationary-phase evaluation 
within the semiclassical method; see Sec.~\ref{sec:DOS_semiclassical-GFs}).
For $E>0$, these terms can be extracted by performing the integral in the complex plane 
with a counterclockwise contour comprised of the half real axis $L_{1}$ ($ {\rm Re} ({ z }) \in [0,\infty)$),
the upper-half imaginary axis ${L}_{2}$ ($ {\rm Im} ({ z }) \in [0,\infty)$) 
traversed downward circumventing the poles, 
  and a circular path $C_{R}$ of large radius $R \rightarrow \infty$ connecting them.
As the integrand has no singularities in the first quadrant (excluding the boundary poles on the imaginary line),
  and as $\int_{C_{R}} F({ z }) d { z } = 0$ in the limit $R\rightarrow \infty$, it follows that
  the original integral~(\ref{eq:dos_S-integral_02}) along the half real axis $L_{1}$
  can be computed with the integral $\int_{L_2} F({ z }) d { z }  $
  (traversed backward up and including the poles along the path itself).
  Thus, with $\alpha= 1/\omega$, Eq.~(\ref{eq:dos_S-integral_02}) gives
\begin{equation}
 \rho^{(S)}(E) 
 =
        \frac{\alpha}{2\pi \hbar} 
\;{\rm Re} \; 
\left[
\pi i 
\left( \frac{R_{0}}{2} + \sum_{k=1}^{\infty} R_{k} \right)
\right]
+ f(E)
=
  \frac{\alpha}{2 \hbar} 
\;{\rm Im} \; 
\left(
e^{\pi \zeta} + 1 
\right)^{-1}
+ f(E)
      \; ,
      \label{eq:dos_S-integral_sc-correction_01}
\end{equation}
where each pole ${ z }_{k} = i \pi k$,
with $k > 0$, yields half the usual residue amount 
(due to a circumventing half-circle), while $k=0$ gives only a quarter of the usual amount.
It is noteworthy that, for $E<0$, a similar procedure involves a clockwise contour closed 
 with an infinite circular path in the lower-half plane; and 
  ${L}_{2}$ along the lower-half imaginary axis going through similar poles 
  ${ z }_{k} = i \pi k$, with
  $k=0, -1, -2, \ldots$:
  the outcome is a reversal in the sign of the energy, i.e., 
  this conforms to the even nature of the density of states that can be deduced from
 Eq.~(\ref{eq:dos_S-integral_02}).
The resulting value of the density of states, Eq.~(\ref{eq:dos_S-integral_sc-correction_01}), for $E>0$,
 consists of: 
(i) the pole contributions, forming a geometric series (with the displayed sum), which reproduce
 the semiclassical correction $\Delta \rho_{\rm sc}$ of Sec.~\ref{sec:DOS_semiclassical-GFs};
 and (ii)
 an extra term $f(E)$, which is the Cauchy principal value of the integral
 along the imaginary axis, and includes the average density plus higher orders in the expansion.
 An interesting feature of Eq.~(\ref{eq:dos_S-integral_sc-correction_01}) 
 is that it involves an infinite geometric series with ratio 
 $-  e^{- \pi \zeta} = -  e^{-i \pi \mu} e^{- \beta E} $, which is 
proportional to an effective Boltzmann factor $e^{- \beta E}$ at the diamond temperature,
Eq.~(\ref{eq:diamond-temperature}),
as follows by identification of the inverse temperature $\beta = \pi \alpha$.
These findings are further discussed in the next section,
 in the context of a systematic semiclassical approximation.

 \subsection{Semiclassical framework and Gutzwiller trace formula}
\label{sec:DOS_semiclassical-GFs}

A general framework for the computation of the density of states---and more generally of all the functional integrals---is
available within the semiclassical approximation.
In the Gutzwiller approach~\cite{Gutzwiller_GTF,Gutzwiller_Chaos-ClassQM},
which applies to the class of Hamiltonians of the form $p^2/2+V({\bf r})$,
this is obtained via the asymptotic expansion of the path integral with respect to $\hbar$, leading to
 the semiclassical energy Green's function 
 \begin{equation}
G({\bf r}'',{\bf r}';E)
=
\overline{G} ({\bf r}'',{\bf r}';E)
+ \sum_{\gamma} A_{\gamma} ({\bf r}'',{\bf r}';E)
\, \exp{\left[\frac{i}{\hbar}
\left(W_{\gamma} ({\bf r}'',{\bf r}';E)
- \mu_{\gamma}\frac{\pi}{2}\right)\right]} 
\label{eq:SC-EnergyGreenF}
\; .
\end{equation}
In Eq.~(\ref{eq:SC-EnergyGreenF}),  
the sum runs over all classical trajectories $\gamma$ joining
${\bf r}' $ to $ {\bf r}''$ at given energy $E$ but arbitrary times $T$, 
 $W_{\gamma}$ is the Jacobi action along the trajectory, 
$\mu_{\gamma}$ counts the number of points conjugate to ${\bf r}' $ in energy,
and the amplitude $A_{\gamma} = -\sqrt{D_{\gamma}}/[i\hbar(2\pi i \hbar)^{(d-1)/2}]$ enforces semiclassical probability conservation, with $D_{\gamma} ({\bf r}'',{\bf r}';E)$ being the determinant of a Hessian matrix consisting
of derivatives with respect to ${\bf r}''$, ${\bf r}'$, and $E$~\cite{Kleinert-PI,Grosche-PI}.
In addition, $\overline{G} ({\bf r}'',{\bf r}';E)$ is the contribution from the stationary point at $T=0$
 (singular part as ${\bf r}' \rightarrow {\bf r}''$) that yields the average density of states.
 
This semiclassical Green's function 
  provides not only the density of states, 
 which is a standard tool for quantum chaos and stability analyses~\cite{Haake_QChaos,Stockmann_QChaos},
 but also a more general approach to address all semiclassical questions.
 Careful evaluation of the trace of Eq.~(\ref{eq:SC-EnergyGreenF})
 gives the Gutzwiller trace formula~\cite{Gutzwiller_GTF,Gutzwiller_Chaos-ClassQM}, which reads
[see Eq.~(\ref{eq:dos-expression})
and Refs.~\cite{Kleinert-PI,Grosche-PI}]:
\begin{align}
    \rho^{}(E) 
    =
     -\frac{1}{\pi} \;{\rm Im}
      {\rm Tr}
       \left[
        \hat{G}^{}(E) \right]
         =
      \overline{\rho}(E) 
      +  \Delta \rho_{\rm sc}  
     \; ,
     \label{eq:dos-from-trace}
\end{align}
where the Green's function trace,
 \begin{equation}
  \begin{aligned}
&
    {\rm Tr}
       \left[
        \hat{G}^{}(E)  \right]
        =
       {\rm Tr}
       \left[
        \hat{\overline{G}}^{}(E)  \right]
      \\
      & 
-
\left( \frac{i}{\hbar} 
    \sum_{\gamma_{p}} \sum_{k=1}^{\infty}
    \frac{T_{\gamma_{p}}
     }{ \left|  {\rm det} \, \left(M_{\gamma_{p}}^{k}
      -\mathbb{I} \right) 
      \right|^{1/2}}
    \;
    \exp\left[
    \frac{i}{\hbar}
    k W(E_{\gamma_{p}}) 
    - i \frac{m_{\gamma_{p}, k}\pi}{2}\right]
    \right)
    \; ,
    \label{eq:Green-trace}  
     \end{aligned}
     \end{equation}
 is the critical functional in the microcanonical version of the theory.
 In Eqs.~(\ref{eq:dos-from-trace})--(\ref{eq:Green-trace}), both the
 trace and the density of states are given by the sum over all periodic orbits 
 $\gamma$, including primitive orbits $\gamma_{p}$ and their $k$-fold repetitions $k =1,2, \dots$, 
with $T_\gamma$ being their corresponding periods; and
  \begin{equation}
  W(E_\gamma)= W(E_{\gamma_{p},k})= 
  k W(E_{\gamma_{p}})
  \end{equation}
  is the closed-path Jacobi action.  
In addition, $m_{\gamma}$ is the Maslov index for the trajectory (a modified extension of $\mu_{\gamma}$);
and $M_{\gamma}$ is the monodromy matrix defined through the growth of a perturbation 
around the primitive orbit.
The leading order  $ \overline{\rho}(E) $ is the ``classical'' density or Thomas-Fermi 
term discussed in Sec.~\ref{sec:app_DOS_exact-GFs};
the other terms give an infinite sum: the semiclassical correction $\Delta \rho_{\rm sc}$.

It should also be noted that the one-dimensional case ($d=1$)
involves the usual subtleties associated with the choice of either a one-sided or a two-sided range 
 (whole real line or half the real axis). 
 This choice, for a given potential, involves the original formula~(\ref{eq:Green-trace}) for the 
 unrestricted two-sided case, where the trace covers the whole real line.
  But the formula is reduced with a factor $1/2$ for the restricted one-sided case, when
  the real line is cut off into two disjoint pieces---see the explanation 
  at the end of Appendix~\ref{sec:app_PI-functionals}, Sec.~\ref{sec:app_canonical-PIs},
  and in the paragraph that includes Eq.~(\ref{eq:1D-R-propagator}).
  In this restricted case, the trace covers only half of the whole range---this adjustment 
  does apply to the important case of the operator $S$ below, where the inverse square potential 
  enforces the interval separation. 

In this work, we are specifically interested in the time evolution dynamics generated by the
hyperbolic operator $S$. As it is unbounded, it does not have any primitive orbits,
and Eqs.~(\ref{eq:dos-from-trace})--(\ref{eq:Green-trace}) give no direct information.  
However, the analytic continuation defined by Eq.~(\ref{eq:S-to-R_alpha}) 
 extends it to the operator $R$, which does have the closed orbits
  discussed in Secs.~\ref{subsec:thermality-Jacobi}
  and \ref{subsec:microcanonical};  the results of those sections provide the Jacobi action and
the period, $W(E_{\gamma})$ and $T_{\gamma}$,
 Eqs.~(\ref{eq:action-R})--(\ref{eq:period-from-action_02}) 
 and (\ref{eq:R-period-microcanonical}).
An analogous method has been previously used for the simpler inverted harmonic oscillator problem 
and the $D$ operator of CQM~\cite{majhi-nh-thermality1}.
Moreover, the Green's functions of the Euclideanized operator $R$ give the same answers, 
properly extended, as those for operator $S$, according to
Eq.~(\ref{eq:GF-relation_R-to-S}),
where additional regularization may be in order, as seen in Sec.~\ref{sec:app_regularization-SC-DOS}.
 Now, in the evaluation of Eq.~(\ref{eq:Green-trace})
 for the operator $R$, which involves an effective one-dimensional problem:
for a given energy, there is only one primitive
periodic orbit, with period given
by Eq.~(\ref{eq:R-period-microcanonical})---and this is already
understood to be an analytic continuation, with an imaginary value.
In addition, the Maslov indices are 
$m_{\gamma_{p}, k} =k m_{\gamma_{p}}$,
where the primitive Maslov index is
$m_{\gamma_{p}}=2$; and the monodromy matrix gives a trivial constant factor equal to one.
In addition, as discussed above, the trace for a one-sided dimensional problem gets reduced by a factor 1/2. 
Then,
 from Eqs.~(\ref{eq:action-R})--(\ref{eq:action-R_Langer-corrected}), which read
$W(E_{\gamma}) \rightarrow \left. W(E_{\gamma}) \right|_{\rm Langer-corr}
 = \pi \alpha_{\!_{R}}\,E_{\gamma} - \, \pi \mu  \, \hbar$
 with the Langer correction in standard units, the exponent in Eq.~(\ref{eq:Green-trace}),
 with $\alpha_{\!_{R}} = i\alpha_{\!_{S}} = i \alpha$,
  becomes $ik \left[ i \pi \alpha E/\hbar - \pi \mu - \pi \right] $, 
and the analytic continuation of the density of states of Eq.~(\ref{eq:dos-from-trace}) takes the form 
\begin{equation}
\begin{aligned}
    \rho^{(S)}(E) 
    & =  
      \overline{\rho}^{(S)}(E) +
    {\rm Im} \left( \frac{i}{2\pi\hbar} 
( i \pi \alpha)
\,
  \sum_{k=1}^{\infty}
    \;
    e^{-ik \pi}
    \,
    \exp\left[
    k 
    \left(
-  \pi \alpha \frac{E}{\hbar}
         - i \pi \mu
    \right)
    \right]
    \right)
    \\
    &
    =
          \overline{\rho}^{(S)}(E) 
          -
  \frac{\alpha}{ 2 \hbar}
  \, 
      {\rm Im}
      \left[
        \sum_{k=1}^{\infty}
    \;
    \left(
-   e^{-i \pi \mu} \,  e^{- \pi \alpha E/\hbar}
\right)^k
\right]
    \; .
 \label{eq:dos-GTF_R-operator}
\end{aligned}
\end{equation}
The semiclassical density of states of Eq.~(\ref{eq:dos-GTF_R-operator})
consists of the Thomas-Fermi  approximation average plus the semiclassical correction 
$\Delta \rho_{\rm sc}$~\cite{Kleinert-PI}, which is identical to the corresponding terms 
in the pole-induced contribution of Eq.~(\ref{eq:dos_S-integral_sc-correction_01}) to the exact density of states.
This correction $\Delta \rho_{\rm sc}$ consists of
 the $k$-fold repetitions that give a geometric series with ratio proportional to the Boltzmann factor $e^{- \beta E}$
at the diamond temperature, as follows by identification of the inverse temperature.
The apparent or effective Boltzmann factor is subject to the following qualifications:
(i) strictly speaking, a temperature dependence of a density of states is not a fully meaningful concept; (ii)
the Boltzmann factor can be considered a sort of precursor or signal of the 
underlying thermal behavior in the form of thermal fluctuations in a bath at temperature $T$;
(iii) the expressions have to be properly regularized.
But, most importantly, in a rigorous treatment of the system,
the complete thermal behavior should be properly probed and identified with the canonical partition 
function, as in Eqs.~(\ref{eq:partition-function-S_01-b}) and (\ref{eq:propagator-trace-relation_explicit}),
or with the density matrix, as in Eq.~(\ref{eq:partition-function-S_03}).
Indeed, an inverse Fourier transform of the terms in 
Eq.~(\ref{eq:dos-GTF_R-operator}) yields a sum with simple poles at 
$T= i \pi \alpha k$, with $k$ integer, which, by Mittag-Leffler's theorem~\cite{Ahlfors_complex-analysis},
corresponds to the partition function~(\ref{eq:partition-function-S_01-b}).
It should be noted that, even though this is a one-dimensional problem,
for which some aspects of the use of the Gutzwiller framework become trivial,
an inverted oscillator generates a dynamic instability that still appears to lead to partially insightful information
via the trace formula, Eqs.~(\ref{eq:dos-from-trace})--(\ref{eq:Green-trace}).

\end{appendix}

\end{document}